\documentclass[prX,superscriptaddress,preprint,11pt]{revtex4}
\usepackage[utf8]{inputenc}
\usepackage{amsmath}
\usepackage{amssymb}
\usepackage{natbib}
\usepackage{graphicx}
\usepackage{color}

\begin{document}
\title{Large phosphorene in-plane contraction induced by interlayer interactions in graphene-phosphorene heterostructures}
\author{Benoit Van Troeye}
\email{benoit.vantroeye@uclouvain.be}
\affiliation{Institute of Condensed Matter and Nanosciences, Universit\'{e} catholique de Louvain, Chemin des étoiles 8, B-1348 Louvain-la-Neuve, Belgium}

\author{Aur\'{e}lien Lherbier}
\affiliation{Institute of Condensed Matter and Nanosciences, Universit\'{e} catholique de Louvain, Chemin des étoiles 8, B-1348 Louvain-la-Neuve, Belgium}

\author{Jean-Christophe Charlier}
\affiliation{Institute of Condensed Matter and Nanosciences, Universit\'{e} catholique de Louvain, Chemin des étoiles 8, B-1348 Louvain-la-Neuve, Belgium}

\author{Xavier Gonze}
\affiliation{Institute of Condensed Matter and Nanosciences, Universit\'{e} catholique de Louvain, Chemin des étoiles 8, B-1348 Louvain-la-Neuve, Belgium}

\begin{abstract}


 Intralayer deformation in van der Waals (vdW) heterostructures is generally assumed to be negligible due to the weak nature of the interactions between the layers, especially when the interfaces are found
 incoherent. In the present work, graphene-phosphorene vdW-heterostructures are investigated with the Density Functional Theory (DFT). The challenge of treating nearly incommensurate 
 (very large) supercell in DFT is bypassed by considering different energetic quantities in the grand canonical ensemble, alternative to the formation energy, in order to take into account the mismatch elastic contribution 
 of the different layers. In the investigated heterostructures, it is found that phosphorene contracts by $\sim$ 4$\%$ in the armchair direction
when compared to its free-standing form. This large contraction leads to important changes in term of electronic properties, 
with the direct electronic optical transition of phosphorene becoming indirect in specific vdW-heterostructures.
More generally, such a contraction indicates strong substrate effects in supported or encapsulated phosphorene -neglected hitherto- and paves the way to substrate-controlled 
stress-tronic in such 2D crystal. 
In addition, the stability of these vdW-heterostructures are investigated as a function of the rotation angle between the 
layers and as a function of the stacking composition. 
The alignment of the specific crystalline directions of graphene and phosphorene is found energetically favored.
In parallel, several several models based on DFT-estimated quantities are presented; they allow notably a better understanding of the global mutual accommodation of 2D materials in their
corresponding interfaces, that is predicted to be non-negligible even in the case of incommensurate interfaces.

\end{abstract}
\maketitle

While the range of properties accessible by 2D materials is already quite broad~\cite{Novoselov2016}, 
the construction of vdW-heterostructures~\cite{Geim2013,Novoselov2016}, 
obtained by stacking different 2D materials on top of the others, enables even more tunability with the addition of the ``out-of-plane'' building freedom. 
In such vdW-heterostructures, the constituent layers are bound to the other ones by weak dispersive forces~\cite{Woods2016}, 
which results in strongly-anisotropic properties, desirable e.g. for electro-optic~\cite{Zhang2016,Ross2017,Latini2017} or battery applications~\cite{Sun2015,Guo2015}. 
For the latter, nanocomposites based on alternating multilayers of graphene and phosphorene -a recently-discovered 2D material~\cite{Liu2014,Castellanos2015}- have demonstrated to outperform graphite
as anode for sodium-ion battery in term of specific capacity 
and cyclability~\cite{Sun2015}. However, understanding and predicting the phase formed by such
vdW-heterostructures, with the underlying questions
of interface coherency~\cite{Porter2009} and commensurability, is still a hot topic, both from experimental and theoretical points of view. 

Indeed, while an exactly-matched graphene/h-BN vdW-heterostructure has been grown recently by molecular beam epitaxy~\cite{Davies2017}, such heterostructure
has more generally been constructed using the pick-and-lift technique~\cite{Novoselov2016}, and then reported either as a mesoscopic commensurate (semi-coherent) 
structure~\cite{Woods2014,Wang2016,Woods2016b} or an incommensurate (incoherent) one~\cite{Woods2014}, depending on the rotation angle
between the layers. 
In the semi-coherent case, areas where graphene is exactly commensurate with h-BN~\cite{Woods2014} are separated by Frenkel-Kontorova domain walls where the accumulated strain is released. 
This leads to regions of preferential and non-preferential stackings, which induces local buckling (corrugation)~\cite{Argentero2017}. 
Such surface reconstruction has also been reported for MoSe$_2$/WSe$_2$ vdW-heterostructure~\cite{Wilson2017}, or for silicene grown on MoS$_2$~\cite{Chiappe2014}.
Even in the case where the lattice mismatch is too large to observe a coherent interface, it is found that the alignment of common crystalline directions is more favorable, like
for MoS$_2$ grown on graphene~\cite{Liu2016} or for germanene grown on MoS$_2$~\cite{Zhang2016b}. For the latter, an important lattice contraction of germanene compared
to its free-standing form has been observed experimentally ($\sim$5\%), indicating that the interlayer interactions play
an important role even in the incoherent phase. 
Although the alignment of crystalline directions is reported as more favorable, other orientations can also be observed experimentally~\cite{Liu2016,Koren2016,Kang2013,Pierucci2016}.

DFT~\cite{Martin2004} could bring valuable informations on this question of interlayer coherence and the underlying question of commensurability/incommensurability of the phase.
Unfortunately, the number of atoms reported experimentally for the primitive cell of semi-coherent vdW-heterostructures (i.e. thousands of atoms for graphene/h-BN) 
limits strongly its use. This problem has been generally tackled by the use of small unit cells associated with large deformations of the involved layers~\cite{Padilha2015},
or by the use of rotation angles that limit the intralayer deformations and the number of atoms to be considered~\cite{Wei2013,Wei2015}. The importance of the errors made by using these approximations 
is generally not investigated, since usual DFT codes are unfortunately unable to do so. 
One has to move to linear-scaling algorithms~\cite{Hine2009,Bell2015,Constantinescu2015,Constantinescu2016} or molecular dynamics simulations~\cite{Slotman2015} to overcome this difficulty.
Still, exploring the whole space of degrees of freedom (rotation angle and translation between the layers, as well as intralayer deformations) is an incredibly challenging task~\cite{Constantinescu2015}, 
even for these approaches, and generally goes beyond the precision of the technique.


In the present work, we investigate the properties of graphene-phosphorene vdW-heterostructures as 
a function of their composition and stacking arrangement, in view of their application as anode in sodium-ion battery~\cite{Sun2015}. 
Thanks to DFT, we explore the energetics and structural properties of a series of small and medium unit
cells, but also extract more transferable thermodynamic quantities, and perform additional analyses
based on a simple energy model, in order to alleviate to a large extent our DFT computational limitations. 

The graphene-phosphorene vdW-heterostructures studied in this work consist of periodically-repeated out-of-plane 
few-layers-thick graphene and phosphorene domains, bound to another by weak dispersive forces.
A special emphasis focuses on scanning both the rotational-angle and the intralayer-deformation degrees of freedom. For the first one, we investigate different commensurate structures and compare their energies
with respect to different definitions of chemical potentials. The errors made by working with commensurate structures on energetics,
can be greatly reduced by comparing the vdW-heterostructures to strained graphite and black phosphorus with the corresponding definitions of modified thermodynamical quantities.
Furthermore, we find that the most favorable angular configuration corresponds to the alignment of the crystalline directions of graphene
and phosphorene (i.e. zigzag on zigzag and armchair on armchair). When such an aligned angular configuration is realized, all stacking and composition arrangements are found nearly equivalent in energy, 
indicating only very small energy penalty to form 
the graphene-phoshorene vdW-interface compared to graphene-graphene and phosphorene-phosphorene vdW-interfaces. 

Concerning now the intralayer deformation, a qualitative model based on ab-initio computed quantities and on the estimation of vdW energy thanks to Grimme's DFT-D3 method~\cite{Grimme2010} has been developed. 
This model is validated on the phosphorene/black phosphorus system, where it allows us to understand 
the compression of the phosphorene armchair lattice parameter from the monolayer to the bulk, as observed for example in Ref.~\onlinecite{Constantinescu2016}.
When applied to the graphene-phosphorene vdW-heterostructures, this model reveals that the phosphorene layers compress quite significantly ($\sim 4\%$) 
in the armchair direction in order to accommodate to the graphene lattice, and that even without considering coherency. 
This contraction leads to important modifications in the electronic band structures of the graphene-phosphorene vdW-heterostructure when compared to the ones 
of isolated monolayers (and corresponding multi-layers), turning the direct electronic transition of phosphorene into an indirect one. Interestingly, this model indicates that such compressive tendency is unrelated
to the coherency of the phase, and thus appears as well in incommensurate structures. It comes from the direct relation between the van der Waals energy and the in-plane atomic density: increasing the atomic density 
(without changing the out-of-plane distance) lowers the van der Waals energy, at nearly no elastic energy cost for the phosphorene armchair direction.

This paper is organized as follows: first, in Sec.~\ref{BuldingBlocks}, the computational details are presented, including convergence parameters, functional, and analysis of the building blocks (graphene
and phosphorene) used to construct
the vdW-heterostructures. We also detail the methodology followed to build them. 
Second, in Sec.~\ref{Sec2}, the different commensurate vdW-heterostructures are characterized and their stabilities as functions of the rotation angle between the layers, composition and stacking arrangement are discussed.
In Sec.~\ref{Sec3}, a structural model, whose purpose is to investigate the intralayer-deformation degrees of freedom, is proposed, its validity verified on graphite and black phosphorus, 
and then applied to the specific case
of the graphene-phosphorene vdW-heterostructures.
Finally, in Sec.~\ref{Sec4}, the electronic properties of the graphene-phosphorene vdW-heterostructures are presented and discussed.
The Supplementary Materials~\footnote{See Supplemental Material \textit{link} for additional figures related to the question of lattice conformation, translational degree of freedom  and of Moiré interference pattern.}
include additional figures, not presented in the main manuscript for sake of readability.

\section{Methodology \& Bulding blocks} \label{BuldingBlocks}

All computations are performed using the \textsc{Abinit} software \cite{Abinit2005,Abinit2009,Gonze2016}. The exchange-correlation energy is approximated using the GGA-PBE functional~\cite{Perdew1996}, corrected
by Grimme's DFT-D3 for the long-range e$^-$-e$^-$ correlation~\cite{Grimme2010,VanTroeye2016}. For the sake of brevity, this combination of exchange-correlation and dispersion corrections 
will be denoted as PBE-D3 in the following.
The cut-off radius for the coordination number, required for the dispersion corrections, is set 
to 105~\AA $\,$  and only pairs contributing for more than $10^{-12}$ Ha are taken into account. 
Calculations are based on plane-waves and ONCVPSP norm-conserving pseudopotentials \cite{Hamann2013} from the PseudoDojo project \cite{VanSetten2017} thus including
multiple angular projectors. A planewave energy cut-off of 42~Ha and a 18$\times$18$\times$1 Monkhorst-Pack wavevector grid~\cite{Monkhorst1976}
are found sufficient for convergences of the ground-state properties of graphene and phosphorene building blocks. The in-plane wavevector mesh is then adapted accordingly to the 
size of the supercell used to build the vdW-heterostructures. A 8 wavevector out-of-plane sampling is used for their bulk counterparts (bernal graphite, 
black phosphorus and the graphene-phosphorene vdW-heterostructures).
A Gaussian smearing of 0.01 Ha is applied for the occupation of states~\cite{Methfessel1989}.  
For the computation of isolated monolayers, cells with a 30 Bohr out-of-plane lattice vector are used. 

This combination of exchange-correlation, pseudopotentials and dispersion corrections 
yields both reasonable lattice parameters and cohesive energies when compared to experiments or Diffusion Monte-Carlo (DMC)
in the case of graphite and black phosphorus as shown in Tab.~\ref{Tab1} and Refs.~\cite{VanTroeye2016,Shulenburger2015} for more comparisons. 

Note still that properly describing the ground-state properties of black phosphorus is a challenging task. Indeed, e.g. none of the usual dispersion corrections or vdW functionals in the DFT framework (DFT-D, vdW-DF
or one-shot TS-vdW)
are able to properly describe the change in the electronic density of the constituent layers going from phosphorene to black phosphorus as predicted by high-order methods~\cite{Shulenburger2015}.
Nonetheless, as the properties of interest are mostly structural and energetical, the chosen combination is a reasonably good approximation for the present study.

The interlayer distance $d_{\text{int}}$ is defined as the distance between the top of a layer and the bottom of the adjacent one.
PBE-D3 yields 3.19~\AA \, and 3.48~\AA, for black phosphorus and graphite, respectively. The predicted intrinsic thickness of phosphorene $h^\text{P}$ is 2.12~\AA \,
in its free-standing form and 2.14~\AA \, in black phosphorus. The latter value compares relatively well with its corresponding experimental counterpart (2.17~\AA~\cite{Cartz1979}). 


\begin{table}[h]
\caption{\label{Tab1} Zigzag lattice parameters $a$, armchair lattice parameters $b$, out-of-plane lattice parameters $c$, cohesive energies and elastic constants of black phosphorus and graphite as
predicted by PBE-D3. Corresponding experimental results~\cite{Cartz1979,Baskin1955,Zacharia2004,Kozuki1991}, as well as other theoretical results~\cite{Appalakondaiah2012,Bucko2013,Shulenburger2015,Savini2011} are mentioned for comparison.}
\begin{tabular}{l c c c | c c}
 \hline \hline
& \multicolumn{3}{c| }{Black phosphorus} & \multicolumn{2}{c}{Graphite} \\ 
\cline{2-6}

 & \multicolumn{5}{c}{Lattice parameters [\AA]} \\
\cline{2-6}
 & $a^\text{P}$ & $b^\text{P}$ & $c^\text{P}$ & $a^\text{C}$ & $c^\text{C}$\\
\cline{2-6}
PBE-D3 & 3.31 & 4.42 & 10.7  & 2.46 & 6.97  \\
TS-vdW~\cite{Appalakondaiah2012,Bucko2013} & 3.29 & 4.39 & 10.82 
 & 2.46 & 6.68 \\
Exp.~\cite{Cartz1979,Baskin1955} & 3.3133 & 4.374 & 10.473 & 2.4589 & 6.7076 \\
\hline
 & \multicolumn{5}{c}{Cohesive energy [meV/atom]} \\
 \cline{2-6}
 PBE-D3 &  \multicolumn{3}{c|}{-91.4} & \multicolumn{2}{c}{-49.1} \\
 TS-vdW~\cite{Shulenburger2015,Bucko2013} & \multicolumn{3}{c|}{-95} & \multicolumn{2}{c}{-55} \\
 DMC~\cite{Shulenburger2015}  &\multicolumn{3}{c|}{-81$\pm$ 6}  & \\ 
 Exp.~\cite{Zacharia2004}  & & &   &\multicolumn{2}{c}{-52 $\pm$ 5} \\ 
 \hline
 & \multicolumn{5}{c}{Elastic constants [GPa]} \\
  \cline{2-6}
 & $c^\text{P}_{11}$ & $c^\text{P}_{22}$  & $c^\text{P}_{33}$ & $c^\text{C}_{11}$ & $c^\text{C}_{33}$ \\
 PBE-D3 & 184 & 39.7 & 40.8 & 1035 & 25.0 \\
 TS-vdW~\cite{Appalakondaiah2012} & 185.9 & 36.8 & 30.6  & & \\
 LDA~\cite{Savini2011} && & & 1109 & 29 \\
 Exp.~\cite{Kozuki1991,Bosak2007} & 178.6 & 55.1 & 53.6 & 1109$\pm$16 & 38.7$\pm$7 \\
 Exp.~\cite{Yoshizawa1986} &  284 & 80 & 57 \\
 \cline{2-6}
 & $c^\text{P}_{12}$ & $c^\text{P}_{23}$  & $c^\text{P}_{13}$ & $c^\text{C}_{12}$ & $c^\text{C}_{13}$ \\
 \cline{2-6}
 PBE-D3 & 35.2 & -3.06 & 2.35 & 195 & -2.54 \\
  TS-vdW~\cite{Appalakondaiah2012} & 16.2 & -0.9 & -0.6 & & \\
   LDA~\cite{Savini2011} && & &175& -2.5 \\
  Exp.~\cite{Bosak2007} &  && & 139$\pm$ 36 & 0$\pm$ 3 \\

  \cline{2-6}
  & $c^\text{P}_{66}$ & $c^\text{P}_{55}$  & $c^\text{P}_{44}$ & $c^\text{C}_{66}$ & $c^\text{C}_{44}$ \\
 \cline{2-6}
  PBE-D3 & 56.4 & 11.9 & 4.48 & 420 & 2.48 \\
  TS-vdW~\cite{Appalakondaiah2012} & 57.4 & 31.2 & 23.4 & & \\
  LDA~\cite{Savini2011} & & & & 467& 4.5 \\
  Exp.~\cite{Kozuki1991,Bosak2007} & 14.5 & 11.1 & 5.5 & 485$\pm$ 10 &5.0$\pm$ 3  \\
  Exp.~\cite{Kozuki1991} & 15.6 & 21.3 &  \\
   Exp.~\cite{Yoshizawa1986} &  59.4 & 17.2 & 10.8 \\
\hline \hline
\end{tabular}

\end{table}

In Tab.~\ref{Tab1}, the elastic constants of these materials, that have been computed using the Density Functional Perturbation Theory (DFPT)~\cite{Hamann2005,Hamann2005b,Hamann2005c, VanTroeye2017} at the relaxed geometries,
are compared to experiments and other theoretical works. To reach 1 GPa precision,
a planewave cut-off energy of 54.5 Ha is required, as well as 24$\times$24$\times$12 and 14$\times$14$\times$14 Monkhorst-Pack grids for graphite and black phosphorus, respectively. The elastic constants 
of graphite are relatively well reproduced in term of relative errors, taking into account the overestimation of the theoretical volume~\footnote{Indeed, $c_{\alpha\beta} = \Omega_0^{-1} \partial^2E/(\partial\epsilon_{\alpha}
\partial\epsilon_{\beta})$, where $\Omega_0$ is the optimized primitive cell volume.} and the experimental error bars. Correcting the volume, they are comparable as well to the ones obtained with LDA~\cite{Savini2011}.
For black phosphorus, the discrepancies between the reported experimental values are quite important~\cite{Kozuki1991,Yoshizawa1986},
varying up to a factor 4 for $c^\text{P}_{66}$ (14.5 GPa in Ref.~\onlinecite{Kozuki1991} compared to 59.4 GPa in Ref.~\onlinecite{Yoshizawa1986})
or a factor 2 for $c^{\text{P}}_{55}$ within the same reference~\onlinecite{Kozuki1991}. Compared to other theoretical methods like one-shot TS-vdW~\cite{Tkatchenko2009,Appalakondaiah2012}, the
PBE-D3 elastic constants differ significantly, except for $c^\text{P}_{11}$, $c^\text{P}_{22}$ and $c^\text{P}_{66}$. Such large variations of elastic constants between functionals and dispersion corrections are also
observed with other functionals and dispersion corrections~\cite{Appalakondaiah2012}.
Thus, with such a large spread of experimental and theoretical values, no conclusive insights can be directly drawn regarding the accuracy of PBE-D3 elastic constants. 

\begin{figure*}[ht]
 \begin{center}
 \includegraphics[width=1\textwidth]{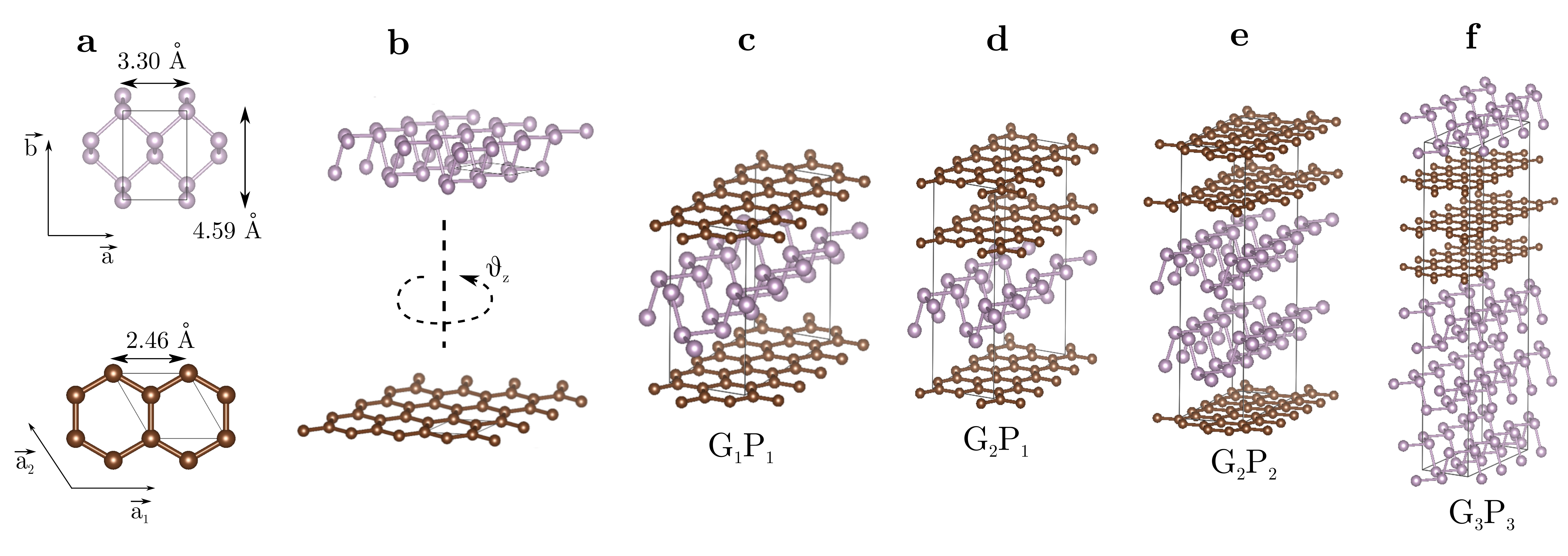}
   \caption{\label{PrimitiveCell}Schematic construction and representation of the graphene-phosphorene vdW-heterostructures studied in this work. (a) Top view of the primitive cells of graphene and phosphorene,
  the two constituent building blocks of the heterostructure. (b) Superlattices of graphene and phosphorene are built and superposed for a given zigzag rotation angle $\theta_z$ between the layers. For that given angle, 
  commensurate structures can be found with a corresponding strain tensor. From these commensurates, periodic graphene-phosphorene vdW-heterostructures are constructed with different composition stackings
  as illustrated in (c)-(f). These structures are denoted G$_i$P$_j$ where i and  j correspond to the number of adjacent layers of graphene and phosphorene, respectively.
  }
\end{center}
\end{figure*}

In order to build the vdW-heterostructures, we start from the (isolated monolayer) graphene and phosphorene primitive cells (see Fig.~\ref{PrimitiveCell}a) and their predicted lattice parameters as reported in Tab.~\ref{Tab1b}
alongside with other theoretical results obtained with other dispersion corrections~\cite{Liu2014,Constantinescu2016}. The intrinsic thickness of phosphorene is also reported in this table.
Note that the armchair lattice parameter of phosphorene is larger ($\sim$~4\%) than its counterpart in black phosphorus. This expansion has also been observed using optB88-vdW~\cite{Constantinescu2016} (2.5 \%) or
PBE corrected by Grimme's DFT-D2~\cite{Qiu2017} (4 \%), 
and thus it can be safely assumed that it is not a spurious effect of the functional.
In fact, as discussed in the following (see Sec.~\ref{Sec3}), this effect results already from a trade-off between interlayer interactions -including vdW ones- and elastic deformation. 
In the same table, the elastic constants per area unit $\tilde{c}_{\alpha\beta}$, which have been computed using DFPT, are also presented. These elastic constants can be converted to hypothetical equivalent bulk elastic constants by multiplying them by 
the corresponding out-of-plane lattice parameters and give values comparable to the bulk elastic constants reported in Tab.~\ref{Tab1}.

\begin{table}[h]
\caption{\label{Tab1b}Zigzag lattice parameter $a$, armchair lattice parameter $b$ and planar elastic constants (per area unit) of phosphorene and graphene as predicted by PBE-D3. For comparison, we report
the theoretical lattice parameters of Refs.~\onlinecite{Liu2014} and~\onlinecite{Constantinescu2016}.}
\begin{tabular}{l c c c | c c}
 \hline \hline
& \multicolumn{3}{c| }{Phosphorene} & \multicolumn{2}{c| }{Graphene} \\ 
\cline{2-6}

 & \multicolumn{5}{c}{Lattice parameters [\AA]} \\
\cline{2-6}
 & $a^\text{P}$ & $b^\text{P}$ & $h^{\text{P}}$ & $a^\text{C}$ & \\
\cline{2-6}
PBE-D3 & 3.30 & 4.59 & 2.12 & 2.46& \\
PBE~\cite{Liu2014} & 3.35 & 4.62& \\
optB88-vdW~\cite{Constantinescu2016} & 3.34 & 4.54 && \\
 \hline
 & \multicolumn{5}{c}{Elastic constants [J/m$^2$]} \\
  \cline{2-6}
 & $\tilde{c}^\text{P}_{11}$ & $\tilde{c}^\text{P}_{22}$ & $\tilde{c}^\text{P}_{12}$& $\tilde{c}^\text{C}_{11}$ & $\tilde{c}^\text{C}_{12}$ \\
 \cline{2-6}
 PBE-D3 & 103 & 23.0 & 17.3 & 356 & 63.4 \\
   \cline{2-6}
 && $\tilde{c}^\text{P}_{66}$ &&& $\tilde{c}^\text{C}_{66}$ \\
 \cline{2-6}
 PBE-D3&&23.2 & && 145 \\

  \cline{2-4}
\hline \hline
\end{tabular}

\end{table}

Afterwards, similarly to what has been done for example in Ref.~\onlinecite{Koda2016}, superlattices are built for each of the two 2D materials in play.
Then, they are compared to each other
for a given zigzag rotation angle $\theta_z$ between the layers (see Fig.~\ref{PrimitiveCell}b). This angle is obtained by comparing zigzag directions in both
phosphorene and graphene. Due to the symmetry of graphene and phosphorene primitive cells, $\theta_z$ take values between 0$^\circ$ and 30$^\circ$.
The reference ($\theta_z=0^{\circ}$) is taken as the configuration where the zigzag direction of phosphorene is aligned with the one of graphene.
Note that, even in this case, graphene and phosphorene lattices do not match, neither in term of symmetries nor in term of lattice parameters. 

By stretching the phosphorene layer by a strain tensor $\underline{\underline{\epsilon}}$~\footnote{Due to the large ratio of elastic constants between graphite and black phosphorus (at least bigger than ten),
we consider in first approximation that phosphorene takes all the deformation.},
close to unity (deviations remain under 10\%), one can construct a large set of commensurate structures (see Fig.~\ref{PrimitiveCell}c), and that for many different rotation angles. 
Afterwards, the structures that are equivalent by symmetry are discarded using \textsc{Pymatgen}'s geometry analyzer~\cite{Ong2013}.
Finally, the structures that minimize the optimum function ``strain $\times$ surface" are fed to the \textsc{Abinit} code.
Note that the strain can break the equivalence
between zigzag directions of the graphene lattices.
Accordingly, the primary zigzag vector $\vec{a}\,^\text{C}$ is defined as the graphene zigzag vector whose angle is the smallest with respect to the phosphorene zigzag vector
(unambiguously defined). The primary armchair vector $\vec{b}^\text{C}$ is then defined as the 
graphene armchair vector that is the closest to the perpendicular to the primary zigzag vector.
Moreover, the strain tensor used to make graphene and phosphorene lattices commensurate with each other might break the perpendicularity of armchair and zigzag vectors in each material, and 
may induce a misalignment of crystalline directions of graphene and phosphorene.
In order to take into account this effect, the deformation angles $\delta_i$ 
and armchair rotation angle $\theta_a$ are defined as follows:
\begin{equation}
 \theta_a = \theta_z - (\delta_{\text{phosphorene}} - \delta_{\text{graphene}}) \approx \theta_z - 2 (\bar{\epsilon}^\text{C}_{\gamma}-\bar{\epsilon}^\text{P}_{\gamma}),
\end{equation}
where $\delta_i$ corresponds to the angle between the armchair and zigzag directions in the considered 2D lattice (graphene or phosphorene) within the vdW-heterostructure, $\bar{\epsilon}^\text{P}_{\gamma}$ and
$\bar{\epsilon}^\text{C}_{\gamma}$ are the (averaged) in-plane shear strains of phosphorene and graphene lattices compared to their free-standing counterparts, respectively.

It has to be reminded that the systems considered in this work are ``bulk'' graphene-phosphorene vdW-heterostructures, obtained by stacking periodically both graphene and phosphorene layers 
(see Fig.~\ref{PrimitiveCell}c).
Still, in contrast to simple layered materials like graphite, not only a ``translational'' ordering is possible (AA, AB, ...), 
but a ``stacking'' ordering as well, as illustrated for the latter case in Fig.~\ref{PrimitiveCell}d. Different composition of graphene and phosphorene layers in the vdW-heterostructures are also considered.
In the following, to differentiate these different structures, G$_\text{i}$P$_\text{j}$ will denote each considered 
graphene-phosphorene vdW-heterostructure, where i and j are respectively the number of consecutive graphene and phosphorene layers.
Only periodic structures where at most 3 identical layers are nearest-neighbors will be considered here, as illustrated in Fig.~\ref{PrimitiveCell}f. 

\section{Interlayer rotation angles and Stability} \label{Sec2}

\begin{figure}[ht]
 \begin{center}
 \includegraphics[width=0.45\textwidth]{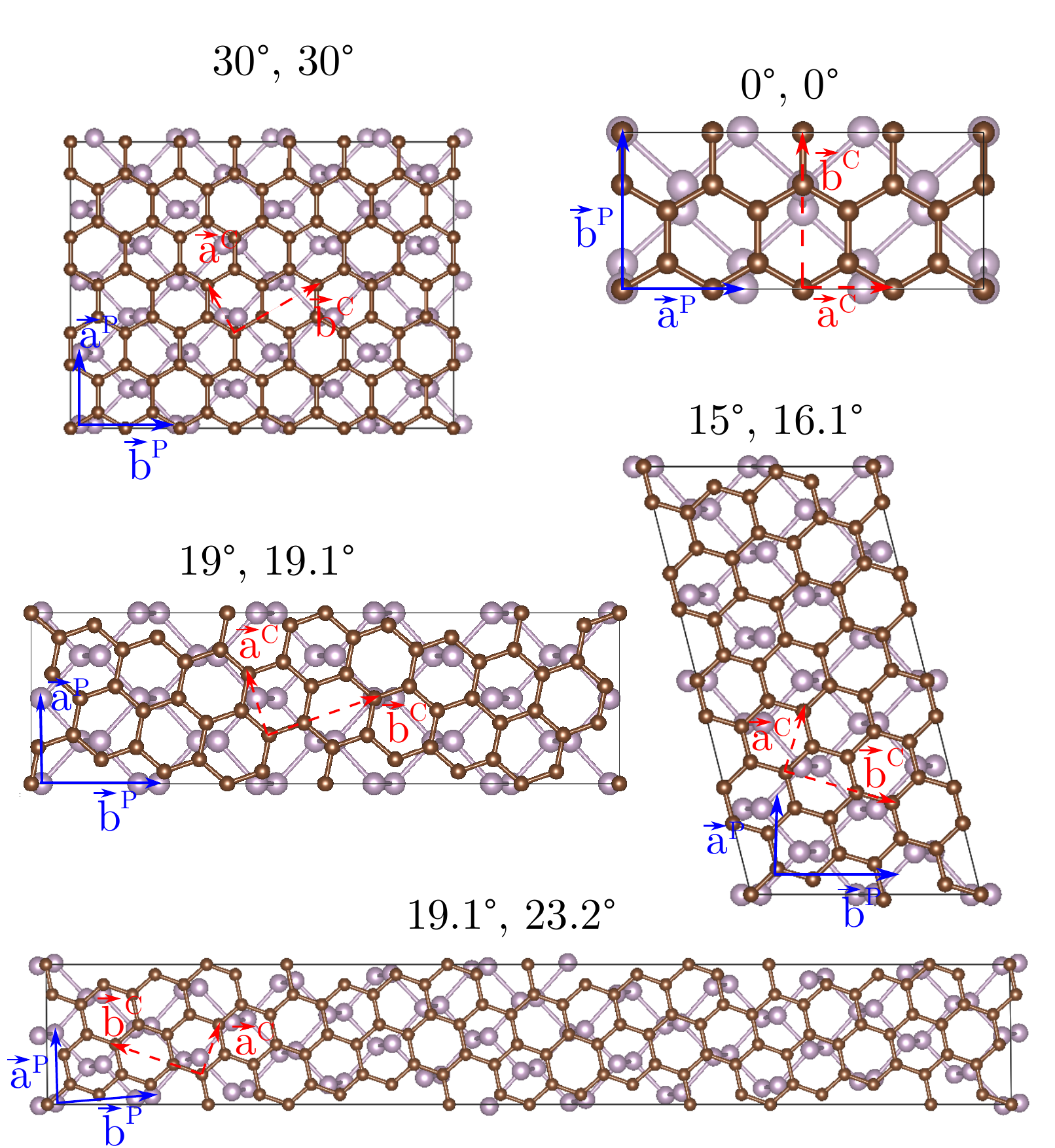}
\caption{\label{Commensurates}The five commensurate structures investigated in this work, characterized here by the zigzag ($\theta_z$) and armchair ($\theta_a$) rotation angles between the graphene and phosphorene layers.
The primary zigzag ($\vec{a}^\text{C}$) and armchair ($\vec{b}^\text{C}$) vectors of graphene are represented in dashed red. Similarly, the primary zigzag ($\vec{a}^\text{P}$) and armchair ($\vec{b}^\text{P}$) vectors of phosphorene
are shown in solid blue. The average strain tensors of graphene ($\underline{\underline{\bar{\epsilon}}}^C$) and of phosphorene
($\underline{\underline{\bar{\epsilon}}}^P$) are then defined based on their respective primary lattice vectors.}
\end{center}
\end{figure}

In this section, the energy landscape of the graphene-phosphorene vdW-heterostructures with respect to the rotation angles between the layers
 is investigated by ab initio means in order to pinpoint  favorable alignments of the constituent layers.
To do so, a small relevant set of commensurate phases was selected among the infinite number of structural configurations for the vdW-heterostructures,
considering the possible effective angles between the layers. 
Here, five different structures consisting of one graphene and one phosphorene layer periodically 
repeated out-of-plane (G$_1$P$_1$) and with a limited number of atoms ($<$200) are investigated. 
These five constructed structures correspond to different zigzag and armchair rotation angles $\theta_z$ and $\theta_a$, and a top view of their primitive cells is shown in Fig.~\ref{Commensurates}.
The numbers of repeated graphene or phosphorene lattice vectors used to build the heterostructure (illustrated in Fig.~\ref{PrimitiveCell}a) are 2D vectors,
denoted respectively $\vec{n}_\alpha^{\text{C}}$ and $\vec{n}_\alpha^{\text{P}}$. The different atomic structures have been optimized in DFT; their characteristics, i.e. rotation angle $\theta_z$ and $\theta_a$,
number of atoms $N_{at}$, atomic fraction of phosphorus atoms $x$
are presented in Tab.~\ref{Tab2}, alongside with the average strain by primitive cell of phosphorene and graphene $\underline{\underline{\bar{\epsilon}}}^\text{P}$ and $\underline{\underline{\bar{\epsilon}}}^\text{C}$.

The first-principle calculations show that graphene distorts only weakly 
to accommodate to phosphorene while the latter takes most of the deformation, 
as expected when one refers to their respective planar elastic constants in Tab.~\ref{Tab1b}. Indeed, the in-plane 
phosphorene elastic constant along the armchair direction $c^\text{P}_{22}$ is 25 times smaller
than the in-plane graphene elastic constant $c^\text{C}_{11}$.
For each of these commensurate structures, the strain remains sufficiently small, such that the change in the intralayer-neighbor distance in
the vdW-heterostructure (distance between neighbors belonging to the same layer) compared to phosphorene/graphene, 
remains similar to the one of black phosphorus/graphite compared to their respective monolayer counterparts, as illustrated in Fig.~S1. For some commensurate structures, 
the strain is not homogeneously shared between all the primitive cells of phosphorene, as illustrated in Fig.~S1c, indicating local deformations.
The average interlayer distance between graphene and phosphorene layers
$\left<d_{\text{int}}\right>$ is reported in Tab.~\ref{Tab3}. The latter drastically varies depending on the effective angle, but always lies in between the graphite (3.48~\AA) and black phosphorus (3.19~\AA) predicted 
interlayer distances. Some corrugations at the graphene-phosphorene interfaces are observed for specific effective angles, and only for phosphorene (graphene stays flat). 
To quantify this effect, the standard deviation of the corrugation height is defined:
\begin{equation}
\begin{split}
  \sigma_d =& \frac{1}{\sqrt{4N_\text{P}}} \left(\sqrt{\sum_{i\in \, \text{top}} (z_t^{\text{P}i} - \bar{z}_t^{\text{P}})^2} \right. \\
  &\left. + \sqrt{\sum_{i\in \, \text{bottom}} (z_b^{\text{P}i} - \bar{z}_b^{\text{P}})^2}\right), \\
\end{split}
\end{equation}
where $z_t^{\text{P}i}$ and $z_b^{\text{P}i}$ are the Cartesian coordinates in the out-plane direction of phosphorus atom $i$ which belongs to the top and bottom planes of phosphorene, respectively;
$\bar{z}_t^{\text{P}}$ and $\bar{z}_b^{\text{P}}$ are the average positions out-of-plane of the top and bottom layers of phosphorene; $N_\text{P}$ is the number of phosphorus atoms in the heterostructure.
This corrugation standard deviation always remains below 0.02 \AA \, and is found to negligible for the $\theta_z=\theta_a=0^\circ$ structure.

  \begin{table*}[ht]{\footnotesize
 \caption{\label{Tab2} Characteristics and in-plane properties of the graphene-phosphorene vdW-heterostructures (vdW-HS) studied in this work. Each vdW-heterostructure is characterized by its
zigzag and armchair rotation angles $\theta_z$ and $\theta_a$, its number of atoms $N_{at}$ and its fraction of phosphorus atom $x$. 
The in-plane properties reported here are the number of graphene and phosphorene primitive cells used along
the heterostructure lattice vectors $\vec{n}^\text{C}_\alpha$ and $\vec{n}^\text{P}_\alpha$ to build the full structure and the averaged graphene and phosphorene strain tensors (in Voigt notation) $\underline{\underline{\epsilon}}^\text{C}$ and
$\underline{\underline{\bar{\epsilon}}}^\text{P}$.}
\begin{tabular}{l | c c c c | c c c c c | c c c c c}
 \hline \hline
vdW-HS & \multicolumn{4}{c |}{Characteristics} &  \multicolumn{5}{c |}{In-plane graphene properties} &  \multicolumn{5}{c}{In-plane phosphorene properties} \\
\cline{2-15}
  & $\theta_z$  & $\theta_a$ & $N_{at}$  & $x$ & $\vec{n}_1^{\text{C}}$($\vec{a}_1^\text{C}$, $\vec{a}_2^{\text{C}}$) &$\vec{n}_2^{\text{C}}$($\vec{a}_1^\text{C}$, $\vec{a}_2^{\text{C}}$)
  & $\bar{\epsilon}^\text{C}_a$& $\bar{\epsilon}^\text{C}_b$ & $\bar{\epsilon}^\text{C}_\alpha$ & $\vec{n}_1^{\text{P}}$($\vec{a}^\text{P}$, $\vec{b}^{\text{P}}$) & $\vec{n}_2^{\text{P}}$($\vec{a}^\text{P}$, $\vec{b}^{\text{P}}$)  
  & $\bar{\epsilon}^\text{P}_a$& $\bar{\epsilon}^\text{P}_b$ & $\bar{\epsilon}^\text{P}_\alpha$  \\
 & [$^\circ$] & [$^\circ$] & [ / ] & [ / ] & [ / , / ] & [ / , / ] &[ \% ]& [ \% ]& [ \% ] &  [ / , / ] & [ / , / ] &  [ \% ]& [ \% ] & [ \% ]\\
 \cline{2-15}
 G$_1$P$_1$($0^\circ$, $0^\circ$) & 0 & 0 & 28 & 0.43 &(4,0) & (1,2)  &0.1 &0.2 & 0.0 &(3,0) &(0,1) & -0.1 & -6.8 & 0.0\\ 
 G$_1$P$_1$($15^\circ$, $16.1^\circ$) & 15 & 16.1 & 96  & 0.42 & (3,-1) & (4,8)&-0.3&0.0 & 0.1 &(0,2)&(5,1) &0.4 & -3.3 &-0.9 \\
 G$_1$P$_1$($19^\circ$, $19.1^\circ$) & 19.0 & 19.1 & 96 & 0.42 &(8,-2) &(2,3)& 0.1& -0.2 & 0.1 &(0,5)&(2,0) & -1.0 & -1.8 & 0.0\\  
 G$_1$P$_1$($19.1^\circ$, $23.2^\circ$) & 19.1 & 23.2 & 192 & 0.42 &(16,-4)&(2,3)& -0.2 & 0.3 &0.5& (1,10)&(2,0)& -1.0 & -1.5 & -3.1\\
G$_1$P$_1$($30^\circ$, $30^\circ$) & 30 & 30 & 148 & 0.43 &(7,0)&(0,3) &0.1&0.8 & 0.0 &(0,4)&(4,0) & -2.1 & -5.9& 0.0 \\
 
 \hline
 G$_1$P$_2$($0^\circ$, $0^\circ$) & 0& 0& 40 & 0.60 &(4,0) & (1,2) &0.1 &0.2& 0.0&(3,0) &(0,1) &-0.1& -6.8&0.0\\ 
 G$_2$P$_1$($0^\circ$, $0^\circ$) & 0& 0& 44 & 0.27 &(4,0) & (1,2) &0.1 &0.2& 0.0&(3,0) &(0,1) &-0.1& -6.8&0.0\\ 
 G$_2$P$_2$($0^\circ$, $0^\circ$) & 0& 0& 56 & 0.43 &(4,0) & (1,2)&0.1 &0.2& 0.0&(3,0) &(0,1)  &-0.1& -6.8&0.0\\
 G$_3$P$_3$($0^\circ$, $0^\circ$) & 0& 0 &84 & 0.43 &(4,0) & (1,2)&0.1 &0.2 &0.0&(3,0) &(0,1) &-0.1& -6.8&0.0\\ 
 \hline \hline
\end{tabular}
}
\end{table*}

\begin{table*}[ht]
\caption{\label{Tab3} Out-of-plane properties of the graphene-phosphorene vdW-heterostructures (vdW-HS) studied in this work i.e.
average interlayer distance $\left<d_{\text{int}}\right>$, 
the standard deviation of the phosphorene corrugation height $\sigma_d$, cohesive energy $E_{\text{coh}}$, mixing energy $E_{\text{mix}}$, stacking energy $E_{\text{stack}}$ and substitution energy $E_{\text{sub}}$.}
\begin{tabular}{l | c c c c c c}
 \hline \hline
vdW-HS  & \multicolumn{6}{c}{Out-of-plane properties} \\
\cline{2-7}
 &
  $\left<d_{\text{int}}\right>$  & $\sigma_d$ & $E_{\text{coh}}$ & $E_{\text{mix}}$&  $E_{\text{stack}}$ & $E_{\text{sub}}$ \\ 
  & [\AA] & [\AA] & [meV/atom] & [meV/atom] &  [meV/atom] & [meV/atom] \\
 \cline{2-7}
 G$_1$P$_1$($0^\circ$, $0^\circ$)  & 3.25 & $\approx$ 0 & -62.8 &4.4 & -68.5& 2.8\\ 
 G$_1$P$_1$($15^\circ$, $16.1^\circ$)& 3.33& 0.005& -61.4 & 5.3 & -62.5 &  4.8 \\
  G$_1$P$_1$($19^\circ$, $19.1^\circ$) & 3.33 &0.01 &-61.5 &5.2 & -62.3 &  4.4\\
 G$_1$P$_1$($19.1^\circ$, $23.2^\circ$) & 3.31 &0.02 & -56.1 &10.6 & -62.3 &4.8\\
 G$_1$P$_1$($30^\circ$, $30^\circ$)  &3.30 & 0.01&-55.1 &12.3 & -65.2&  6.4\\

 \hline
 G$_1$P$_2$($0^\circ$, $0^\circ$)  & 3.23& $\approx$ 0 & -70.4 & 3.8&-78.9& 1.5 \\
 G$_2$P$_1$($0^\circ$, $0^\circ$) & 3.25& $\approx$ 0 & -56.7 & 3.8 &-60.4& 2.6 \\
 G$_2$P$_2$($0^\circ$, $0^\circ$)  & 3.26& $\approx$ 0 & -62.8 & 4.2&-68.8&2.5 \\
 G$_3$P$_3$($0^\circ$, $0^\circ$) & 3.25& $\approx$ 0& -63.5 &3.4 &-69.6&1.6 \\
 \hline \hline
\end{tabular}

\end{table*}

The comparison of the relative stability of these heterostructures can now be performed.
However, the ratio between the number of carbon and phosphorus atoms is not identical in the
different structures, and thus their total energies cannot be straightforwardly compared. 
The standard treatment for comparing materials with varying composition relies on the introduction 
of chemical potentials for each species, in the grand canonical formalism. 
In the present case, however, one can define as reference 
different reservoirs of carbon and phosphorus atoms: either the isolated monolayers (graphene and phosphorene),
or bulk graphite and black phosphorus. Both will be considered,
with different designation: ``cohesive energy" and ``mixing energy".
Later, this analysis of stability will be refined by introducing two other concepts: the ``stacking energy"
and the ``substitution energy".

So, first, let's define the cohesive energy per atom of the vdW-heterostructure with respect to graphene and phosphorene:
\begin{equation}
 E_{\text{coh}} =  \frac{E_{\text{HS}}}{N_{\text{HS}}} - x \, \mu_{\text{P}} - (1-x) \, \mu_{\text{C}},
\end{equation}
where $E_{\text{HS}}$ is the computed total energy of the vdW-heterostructure, $N_{\text{HS}}$ its number of atom per commensurate structure, 
$x$ is the fraction of phosphorus atoms, $\mu_{\text{P}} = E_{\text{P}}/N_{\text{P}}$ the chemical potential of phosphorus computed in phosphorene,
 where $E_{\text{P}}$ is the total energy of phosphorene and $N_{\text{P}}$ its number of atoms per primitive cell. 
This cohesive energy expresses the gain in energy by forming the vdW-heterostructure compared to the isolated layers. 
It is reported in Tab.~\ref{Tab3}.
Note that this definition of cohesive energy for graphite and for black phosphorus is consistent with the one used in Tab.~\ref{Tab1}.
From the negative values in this table, we can conclude that the condensation of phosphorene and graphene layers is always endothermic, thus spontaneous. 
However, it remains to determine which phase condenses preferentially with respect to the others, and in particular whether graphene-phosphorene
heterostructures are stable with respect to graphite and black phosphorus.

Compared to isolated phases of graphite and black phosphorus, a given vdW-heterostructure is more stable if its cohesive energy is smaller than a reference energy
\begin{equation}
 E_{\text{ref}} =  x \, E^{\text{BP}}_{\text{coh}} + (1-x) \, E^{\text{Gr}}_{\text{coh}} 
\end{equation}
where $E^{\text{BP}}_{\text{coh}}$ and $E^{\text{Gr}}_{\text{coh}}$ 
are the cohesive energies of black phosphorus and graphite, respectively. To give an order of magnitude, the reference energy is $-66.5$ meV/atom for $x=0.42$. 
The change of phosphorus fraction between the different heterostructures modifies this last value by les than 1~meV/atom.  

This stability assessment, that differs from the one relative to isolated graphene and phosphorene, is facilitated by the use of the mixing energy instead of the cohesive energy.
The mixing energy expresses the gain (or loss) of energy of the heterostructure compared to graphite and black phosphorus,
\begin{equation}
 E_{\text{mix}} = \frac{E_{\text{HS}}}{N_{\text{HS}}} - x \, \mu^{\text{BP}}_{\text{P}} -  (1-x) \,  \mu^{\text{Gr}}_{\text{C}},\label{Mixing}
\end{equation}
where $\mu^{\text{BP}}_{\text{P}}$ and $\mu^{\text{Gr}}_{\text{C}}$ are the chemical potentials computed in black phosphorus and graphite, respectively. The difference between
the cohesive energy and the mixing energy is simply the reference energy. If the mixing energy is larger than 0, it is less energetically favorable to form the heterostructure than
forming isolated bulk phases of graphite and black phosphorus.
The mixing energies reported in Tab.~\ref{Tab3} are always positive (exothermic), although their magnitudes are quite small so that entropy will probably be
able to drive the mixing at room temperature.

Although this definition of mixing energy allows one to study the stability of different commensurate structures with respect to another, 
it does not allow to investigate directly how the energy varies with respect to the rotation angle between the layers. Indeed, a given commensurate structure may not be the global energy minimum for a given rotation
angle between the layers. In this case, the study of the vdW-heterostructures stability based on a finite set of commensurate structures and as a function of the rotation angles is futile. 
In order to overcome this difficulty, one can first
identify the real energy minimum of a given rotation angle, and then compare it to others in order to identify the most favorable angle. This first solution is investigated using the model presented in
Sec.~\ref{Sec3}. From this model,
one can extract the contributions that need to be added to the mixing energy in order to reach the ground state for a given rotation angle between the layers as discussed latter.
However, one can alternatively try to find a definition of energy that is less sensitive to the choice of the commensurate structures in use.  
Indeed, the relative energy of the different commensurate structures can be split in (1) their difference in elastic energies, needed to match their lattice vectors, 
and (2) their difference in interlayer energies (including van der Waals, but not only) at the commensurate lattice vectors.
As for the previous analysis, one can take as reference either the monolayers or the bulk materials.

First, we define a stacking energy corresponding to the gain in energy obtained by taking as reference isolated graphene and phosphorene layers stretched to match the heterostructure lattice vectors:
\begin{equation}
 E_{\text{stack}} = \frac{E_{\text{HS}}}{N_{\text{HS}}} - x \left.\mu_{\text{P}}\right|_{\underline{\underline{\bar{\epsilon}}}\,^\text{P}_\text{HS}} 
 - (1-x) \left.\mu_{\text{C}}\right|_{\underline{\underline{\bar{\epsilon}}}\,^\text{C}_\text{HS}}.
\end{equation}
In this last expression, $\left.\mu_{\text{P}}\right|_{\underline{\underline{\bar{\epsilon}}}\,^\text{P}_\text{HS}}$ and  $\left.\mu_{\text{P}}\right|_{\underline{\underline{\bar{\epsilon}}}\,^\text{C}_\text{HS}}$
are the chemical potentials of carbon and phosphorus, respectively, computed at the corresponding graphene and phosphorene strained lattices in the vdW-heterostructure $\underline{\underline{\bar{\epsilon}}}\,^\text{P}_\text{HS}$ and
$\underline{\underline{\bar{\epsilon}}}\,^\text{C}_\text{HS}$. These strained lattices correspond
to the ones obtained from the averaged strain tensors reported in Tab.~\ref{Tab2}, and the internal degrees of freedom are optimized.
Compared to the cohesive energy, the stacking energy is less sensitive to the change in intralayer energy due to the strains used to match graphene and phosphorene lattices,
and allows the comparison of direct stacking.

Similarly to this stacking energy, which refers to the free-standing 2D materials, one can define the substitution energy which is based on the chemical potentials computed in the bulk:
\begin{equation}
 E_{\text{sub}} =  \frac{E_{\text{HS}}}{N_{\text{HS}}} - x \left.\mu^{\text{BP}}_{\text{P}}\right|_{\underline{\underline{\bar{\epsilon}}}\,^\text{P}_\text{HS}} - 
 (1-x) \left.\mu^{\text{Gr}}_{\text{C}}\right|_{\underline{\underline{\bar{\epsilon}}}\,^\text{C}_\text{HS}} , \label{Substitution}
\end{equation}
where $\left.\mu^{\text{BP}}_{\text{P}}\right|_{\underline{\underline{\bar{\epsilon}}}\,^\text{P}_\text{HS}}$ and $\left.\mu^{\text{Gr}}_{\text{C}}\right|_{\underline{\underline{\bar{\epsilon}}}\,^\text{P}_\text{HS}}$
are the chemical potentials of phosphorus/carbon computed in black phosphorus and graphite, respectively, and at the corresponding in-plane strain tensors in the vdW-heterostructure. The interlayer distances between
these strained graphene and phosphorene layers and the internal degrees of freedom are optimized. Although relatively simple, this substitution energy for a given rotation angle between
the layers is found to be in good agreement with the results given by the model of Sec.~\ref{Sec3}.
The variation of the substitution energy is only possible through the change of interlayer energies in the vdW-heterostructure, graphite and black phosphorus, which are in principle all different,
but are found to approximatively counterbalance each others.
This definition of substitution energy is thus well-suited in order to investigate the energy landscape as a function of the rotation angle between the layers and to the available literature. 

\begin{figure}[ht]
 \begin{center}
 \hspace*{-1.6cm}\includegraphics[width=0.62\textwidth]{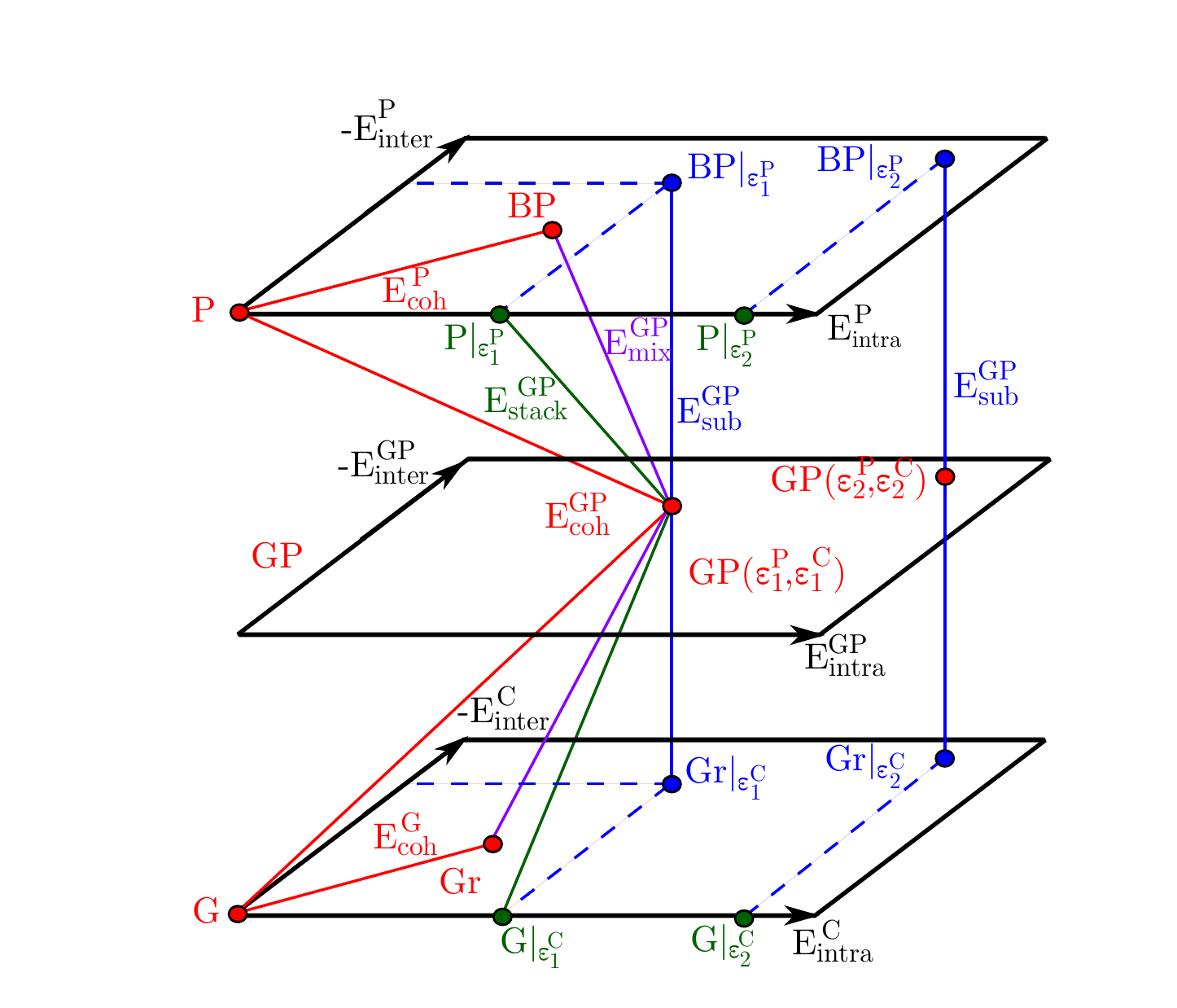} 
 \caption{\label{Scheme}  Schematic representation of the different definitions of energies: cohesive, mixing, stacking and substitution energies. The scheme is split into
  three planes, corresponding each to a different system (upper plane: black phosphorus, middle plane: graphene-phosphorene vdW-heterostructure, bottom plane: graphite). For sake of clarity, the intralayer
  and interlayer energies of each system are the plane axes. G and P abbreviate graphene and phosphorene, respectively, Gr and BP their bulk counterparts (graphite and black phosphorus), while GP stands for 
  the graphene-phosphorene vdW-heterostructure. The cohesive energy $E_{\text{coh}}$ and mixing energy $E_{\text{mix}}$ are trivially defined by the scheme. 
  The stacking energy $E_{\text{stack}}$ is defined from the phosphorene and graphene geometries in 
  the vdW-heterostructure, themselves characterized by the average strain tensors $\underline{\underline{\bar{\epsilon}}}^\text{P}$ and $\underline{\underline{\bar{\epsilon}}}^\text{C}$ (the tensor and average notations are dismissed here for sake of clarity), respectively. Finally,
  the substitution energy $E_{\text{sub}}$ is defined from the same strained phosphorene and graphene in-plane geometries, but now stacked in the optimal stacking sequence on top of each other
  (interlayer distance optimized). By construction, the intralayer energies computed in the vdW-heterostructure and in strained graphite and black phosphorus cancel out. Only remains the difference between
  interlayer interactions. If neglected, the substitution energy is independent of the commensurate structure in use ($E_\text{sub}$[$\epsilon_1^\text{P}$,$\epsilon_1^\text{C}$] =
  $E_\text{sub}$[$\epsilon_2^\text{P}$,$\epsilon_2^\text{C}$]).}
\end{center}
\end{figure}

\begin{figure}[ht]
 \begin{center}
  \hspace*{-0.1cm}\includegraphics[width=0.5\textwidth]{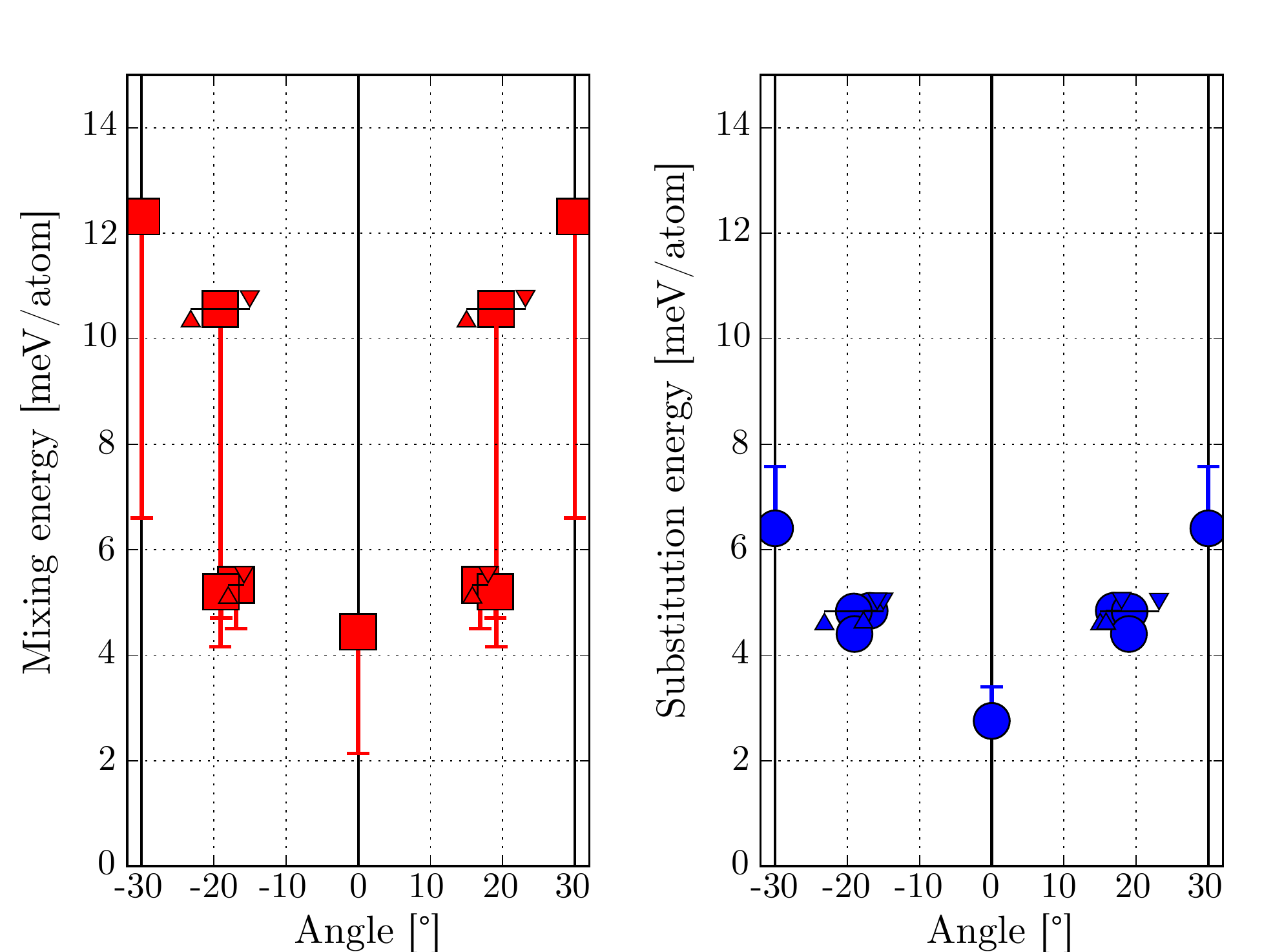}
   \caption{\label{Angle} (Left) Mixing energy $E_{\text{mix}}$  and (Right) substitution energy $E_{\text{sub}}$
  for the different G$_i$P$_j$ vdW-heterostructures as a function of the zigzag and armchair rotation angles $\theta_z$ and $\theta_a$.  The red square and blue circle symbols correspond to the
mixing and substitution energies, respectively, and the angle intervals $[\theta_z,\theta_a]$ are delimited thanks to short horizontal lines with upward and downward arrows. Using the model presented in Sec.~\ref{Sec3},
the differences between a given commensurate structure and the estimated ground state for that specific rotation angle are shown using error bars for both mixing and substitution energies. }
\end{center}
\end{figure}

These different definitions of energy and related chemical potentials are summarized and illustrated in Fig.~\ref{Scheme}, the computed values for the different vdW-heterostructures given in Tab.~\ref{Tab3}, 
and the dependence of mixing and substitution energies with respect to the rotation angles shown in Fig.~\ref{Angle}. 
Using the model presented in Sec.~\ref{Sec3}, the differences between a given commensurate structure and the estimated ground state for that specific rotation angle are shown using error bars, for both the mixing and substitution energies. 

In term of mixing energy, the most favorable structure corresponds to G$_1$P$_1$(0$^\circ$, 0$^\circ$) which is surprisingly not
the one that minimizes the sum of graphene and phosphorene strains (large compression of phosphorene armchair lattice).
The spread in energy between all the investigated structures is about 8 meV/atom. Still, all these structures are quite far ($>$1 meV/atom) from
their corresponding estimated ground states, with the notable exceptions of G$_1$P$_1$(15$^\circ$, 16.1$^\circ$) and G$_1$P$_1$(19$^\circ$, 19.1$^\circ$).
On contrary, the difference in term of substitution energy between the studied structures and their corresponding estimated ground states ($<1.2$ meV/atom) is acceptable in view of the
rotation angle energy landscape. Note that the estimated ground states obtained either through mixing or substitution energies are consistent with respect to another. Based on our results, the substitution
energy appears to be a smooth function of the rotation angle that admits one minimum at $\theta_z=\theta_a=0^\circ$ (zigzag on zigzag and armchair on armchair). The computed spread in substitution energy is of approximatively
3.6 meV/atom, or 4.1 meV/atom using the estimated ground states. Still, based on the stacking energy, the possibility that armchair on zigzag ($\theta_z=\theta_a=$ 30$^\circ$) cannot be excluded and could thus be a favorable alignment as well.

To the best of our knowledge, there have been no theoretical nor experimental investigations on the energy landscape as a function of the rotation angle between the layers for 
graphene-phosphorene vdW-heterostructures. Consequently our results can only be compared to the cases of
other studied vdW-heterostructures, which involve generally much stiffer materials than phosphorene (graphene, h-BN or transition-metal dichalcogenides), with the exception of silicene and germanene.
Theoretical investigations of this rotation degree of freedom~\cite{Constantinescu2015,Woods2016b} have generally been
performed using extremely large supercells, in which the layers kept their original lattice parameters,
sometimes based on experimental observation of lattice matching.
However, this situation should not be hypothesized a priori in ab initio calculations. 
It corresponds to artificially imposing a non-modification of the independent monolayer lattice parameters.
However, for the sake of comparison, it will be assumed here that both their reported energy and
our substitution energies can be directly compared. 

First, similarly to what is found for graphene bilayers~\cite{Koren2016}, graphene on h-BN~\cite{Woods2016b,Wang2016}, silicene on MoS$_2$~\cite{Chiappe2014}, germanene on MoS$_2$~\cite{Zhang2016b} 
and MoS$_2$ on graphene~\cite{Liu2016}
some rotation angles are found more favorable -the so-called magic angles- in the case of the graphene-phosphorene vdW-heterostructure. More specifically,
 the most stable configuration corresponds to the alignment the crystalline directions (e.g. zigzag on zigzag) of
the 2D materials in play in the vdW-heterostructure, similarly to what is reported in Refs.~\onlinecite{Koren2016,Woods2016b,Wang2016,Chiappe2014,Zhang2016b,Liu2016}, although not all of these heterostructures are
reported as commensurate (MoS$_2$ on graphene, for example).
In contrast, Constantinescu and Hine did not find theoretically any remarkably-favorable rotation angle between the layers in
the case of MoSe$_2$ on MoS$_2$~\cite{Constantinescu2015}. Contrarily to what it first seems, we do not think their results go against the previously-mentioned observations. In fact,
we suspect that the energy landscape with respect to rotation is flatter in the case of MoSe$_2$ on MoS$_2$ than in our case due to a) large lattice mismatch b) large elastic constants
for the two layers in play.

Finally, as the most favorable angular configuration has been identified for the graphene-phosphorene vdW-heterostructure as 0$^\circ$,
one can now investigate the stability of these heterostructures as a function of the stacking composition fixing the rotation angles $\theta_z=\theta_a=0^\circ$ (G$_1$P$_1$, G$_2$P$_1$, G$_1$P$_2$, G$_2$P$_2$ and G$_3$P$_3$).
During the relaxation, it is found for these vdW-heterostructures that the most stable translational-stacking configuration is AB-stacking for adjacent graphene layers~\footnote{Translation of 
a carbon-carbon distance in the armchair direction between two adjacent layers}
and for adjacent phosphorene layers~\footnote{Translation of a half zigzag lattice parameter in the zigzag direction between two adjacent layers}. A non-specific translational vector is observed at 
the interface between graphene and phosphorene layers as shown in Fig.~S2 for the G$_1$P$_1$ vdW-heterostructure. 
The cohesive, mixing, stacking and substitution energies of these structures are displayed in Tab.~\ref{Tab2}.
It is found that, by atom, all G$_\text{i}$P$_\text{j}$ configurations are really close in term of substitution energy ($<$1 meV/atom).
This indicates that the formation of a graphene-phosphorene
bulk vdW-composite is not driven by enthalpy, which would lead to specific stacking arrangement like G$_1$P$_1$, but by entropy.
Still, the synthesis process of the graphene-phosphorene bulk vdW-composite could allow one to tune its stacking sequence by, for example, exfoliating and depositing the layers on top of each other
one by one. The experimental stacking composition of vdW-heterostructures is mostly limited by the exfoliation technique. If it produces few-layers thick graphene and phosphorene multi-layers (between 5 and 10)
then the vdW-heterostructures would also be composed of graphene and phosphorene domains of the same thickness.
This may reveal critical for specific applications, when the properties of interest, such as the cyclability in a sodium-ion battery~\cite{Sun2015}, may depend strongly on this stacking composition. 
Finally, it is worth mentioning that the average interlayer distance between graphene and phosphorene layers does not vary significantly with respect to the stacking composition ($\sim 3.25$ \AA), and lies
in between the one of graphite and black phosphorus.



\section{Intralayer deformation} \label{Sec3}

The goal of the present section is to provide an approach to investigate the (eventual) averaged intralayer deformation of the different constituent layers in the graphene-phosphorene vdW-heterostructures.
Indeed, contrary to graphene, h-BN or transition-metal dichalcogenides, phosphorene is soft in the armchair direction, indicating that it is easy to deform it
in order, for example, to maximize its interlayer interactions.
This effect should be relatively decorrelated to the question of phase coherency, 
as it is already observed in black phosphorus. In this material, all the constituent phosphorene layers
match the others by symmetry, independently of the strains they undergo. Still, 
the armchair lattice parameter of phosphorene
varies strongly going from the monolayer to the bulk (see Tabs.~\ref{Tab1} and~\ref{Tab1b}) only due to the interlayer interactions. 
Similar effects are thus expected to appear in the graphene-phosphorene vdW-heterostructures, although
their amplitude has to be determined, which proves to be puzzling.
Indeed, the approach followed in the previous section does not allow to explore the intralayer-deformation degrees of freedom, since working with a given commensurate structure fixes approximatively 
the lattice parameters of phosphorene in order to accommodate to the ones of graphene.
In consequence, the phosphorene-deformation energy map cannot be simply explored without considering different commensurate structures, 
some of them being extremely costly of presently available simulations.
In addition, this intralayer deformation could still occur in the case of incommensurate vdW-heterostructures, as reported
in the case of germanene grown on MoS$_2$~\cite{Zhang2016b}.

In order to overcome this difficulty, a simple model is constructed based on the hypothesis~\cite{Woods2014,Wang2016,Woods2016b} that the important terms that drive surface reconstruction, or intralayer deformation,
are the interlayer (hereby only considering vdW as discussed latter) and intralayer (elastic) energies, and that
they can be decoupled. The general approach is presented here for two given 2D materials in which the crystalline directions are aligned i.e. zigzag on zigzag and armchair on armchair. 
It can be easily extended for any given rotation angle between the layers. 
This model is further verified on graphite and black phosphorus, before being applied to the case of graphene-phosphorene vdW-heterostructure.

The sum of elastic and vdW energies, that determines entirely the deformation of the phosphorene and graphene layers, can be written as
\begin{equation}
 \Delta E  = E_{\text{elast}}(n^{\text{(1)}}_{\alpha},n^{\text{(2)}}_{\alpha}) + E_{\text{vdW}}(n^{\text{(1)}}_{\alpha},n^{\text{(2)}}_{\alpha}), \label{DE}
\end{equation}
where $n^{\text{(1)}}_{\alpha}$ and $n^{\text{(2)}}_{\alpha}$ are the numbers of repeated primitive cells of the first and second 2D materials in the direction $\alpha$ used to build the heterostructure, respectively. 
For sake of simplicity, the effect of shear strains and internal relaxations originating from the counterbalance of vdW forces, interatomic force constants 
and internal strain parameters~\cite{Hamann2005} are neglected. 
Their supercell lattices match:
\begin{equation}
 n^{\text{(1)}}_{\beta} a^\text{(1)}_{\beta} -n^{\text{(2)}}_{\beta} a^\text{(2)}_{\beta} = 0, \label{strain}
\end{equation}
where $a^\text{i}_{\beta}$ are the lattice parameters of the corresponding 2D materials.
The residual strain, resulting from the lattice mismatch between the supercells, is equilibrated between the two layers such as the global surface tension vanishes:
\begin{equation}
 \tilde{\sigma}_{\alpha} = \sum_\beta n^{\text{(1)}}_{\beta}  \tilde{c}^\text{(1)}_{\alpha\beta} \epsilon^{\text{(1)}}_{\beta} + \sum_\beta n^{\text{(2)}}_{\beta}  \tilde{c}^\text{(2)}_{\alpha\beta} \epsilon^{\text{(2)}}_{\beta} = 0, \label{sigma}
\end{equation}
where $\tilde{c}^\text{(1)}_{\alpha\beta}$ and $\tilde{c}^\text{(2)}_{\alpha\beta}$ are the elastic constants of the two 2D materials per surface unit, 
and $\epsilon^{\text{i}}_{\beta}$ their strain by primitive cells. 
With the exception of an equivalent rescaling of the graphene and phosphorene superlattices, the strain is unambiguously defined by the numbers of repeated primitive cells. The total strains undergone by the layers are split
equally between all its constituent cells. To this set of strains correspond a certain elastic energy given by
\begin{equation}
 E_{\text{elast}} = \frac{A^{(1)}_0}{2} \sum_{\alpha,\beta} \tilde{c}^\text{(1)}_{\alpha\beta} \epsilon^{\text{(1)}}_{\alpha} \epsilon^{\text{(1)}}_{\beta} + \frac{A^{(2)}_0}{2} \sum_{\alpha,\beta}
 \tilde{c}^\text{(2)}_{\alpha\beta} \epsilon^{\text{(2)}}_{\alpha} \epsilon^{\text{(2)}}_{\beta}, \label{dEtotal}
\end{equation}
where $A^{(1)}_0$ and $A^{(2)}_0$ are the undeformed surfaces defined by the 2D lattices of the first and second 2D material, respectively. 
Afterwards, the
vdW energy is itself estimated within the DFT-D3 method, that only requires the atomic positions and the exchange-correlation functional, based on the lattice-matched structure. 
The interlayer distance is fixed to the one of graphite and black phosphorus in the case of their respective analysis, and to the one of G$_1$P$_1$(0$^\circ$, 0$^\circ$) in the case of the vdW-heterostructure.
The vdW energy is found nearly independent to the translation between the layers in the case of vdW-heterostructure. 
In all the cases, the vdW energies of the relaxed isolated layers are subtracted to the computed vdW energy (per atom) in order to get only the ``interlayer'' vdW interactions:
\begin{equation}
 E_{\text{vdW}} \rightarrow {E_{\text{vdW}} - x \frac{E^\text{P}_{\text{vdW}}}{N_\text{P}} - (1-x) \frac{E^\text{C}_{\text{vdW}}}{N_\text{C}}},
\end{equation}
where $E^\text{P}_{\text{vdW}}$ and $E^\text{C}_{\text{vdW}}$ are the vdW energies of undeformed phosphorene and graphene, respectively. Afterwards, the reference for the vdW energy is defined as the one in which
the constituent layers are neither stretched nor compressed. 

It is expected that the lattice parameters of the two 2D materials remain relatively similar (less than 10$\%$ at least) in the vdW-heterostructure compared to the ones in their free-standing forms.
In consequence, the ratio between the number of primitive cells of graphene 
and phosphorene in a given direction is chosen as close as the inverse ratio of their free-standing lattice parameters $a_{0,\beta}$,
\begin{equation}
 \frac{n^{\text{(1)}}_{\beta}}{n^{\text{(2)}}_{\beta}} \approx \frac{a^{\text{(2)}}_{0,\beta}}{a^\text{(1)}_{0,\beta}}.
\end{equation}
In practice, $n^{\text{P}}_{1}/n^{\text{C}}_{1}$ ratios range from 7/10 to 4/5 in the zigzag direction 
and $n^{\text{P}}_{2}/n^{\text{C}}_{2}$ ratios ranges from 9/10
to 1/1 in the armchair direction. Interestingly, taking a irrational number for this ratio corresponds in reality to investigate incommensurate superlattices of graphene and phosphorene.
Based on our previous observations, this incommensurate phase corresponds to an unique set of strains, that lies exactly between the ones of two commensurate structures, as an irrational number always lies
between two rational numbers.

Even with this model, scanning the entire configurational space in the case of the vdW-heterostructure is computationally demanding. In consequence, the elastic energy is estimated
for all the combination of superlattice vectors while the vdW energy only for specific points of the mesh. These points, expressed in terms of phosphorene strains, are shown in Fig.~S4.
It is found that the vdW energy in graphite, in black phosphorus and even in the vdW-heterostructure shows on average
a linear dependency with respect to the phosphorene deformation (both along the zigzag and armchair directions). 
Thus, this vdW energy can be interpolated with respect to the zigzag and armchair strains for the remaining superlattices of the graphene-phosphorene vdW-heterostructures. 
Their energy landscape as a function of the phosphorene deformation is illustrated in Fig.~S4.
Hereafter, only the results for specific cuts in the strain planes, which correspond to the lines along which the vdW
energy was estimated, will be shown. 

The fact that both the elastic and vdW energies are varying continuously with strain for phosphorene has an important physical consequence.
Indeed, it implies that even in the case of an incommensurate (incoherent) interface, phosphorene can still accommodate in average to graphene, that the effect can be
quantified and non-negligible, as discussed later in this paper. This effect adds up to the possible local accommodation of phosphorene to graphene, that will determine the type of interfaces 
they form together (i.e. coherent, semi-coherent or incoherent). The Moiré interference pattern of graphene and phosphorene is relatively complex, as shown in Fig.~S3, due to the lack of shared 
symmetries between these 2D materials and their large lattices mismatches. In addition, the Moiré period varies with the averaged strain undergone by the constituent layer (here, phosphorene) in agreement
with what is observed experimentally for graphene grown on h-BN~\cite{Davies2017}. This makes the investigation of local accommodation and its impact on energetics incredibly challenging, and we prefer 
here to highlight already the large effect of a global accommodation of phosphorene to graphene.

\begin{figure}[ht]
 \begin{center}
 \includegraphics[width=0.5\textwidth]{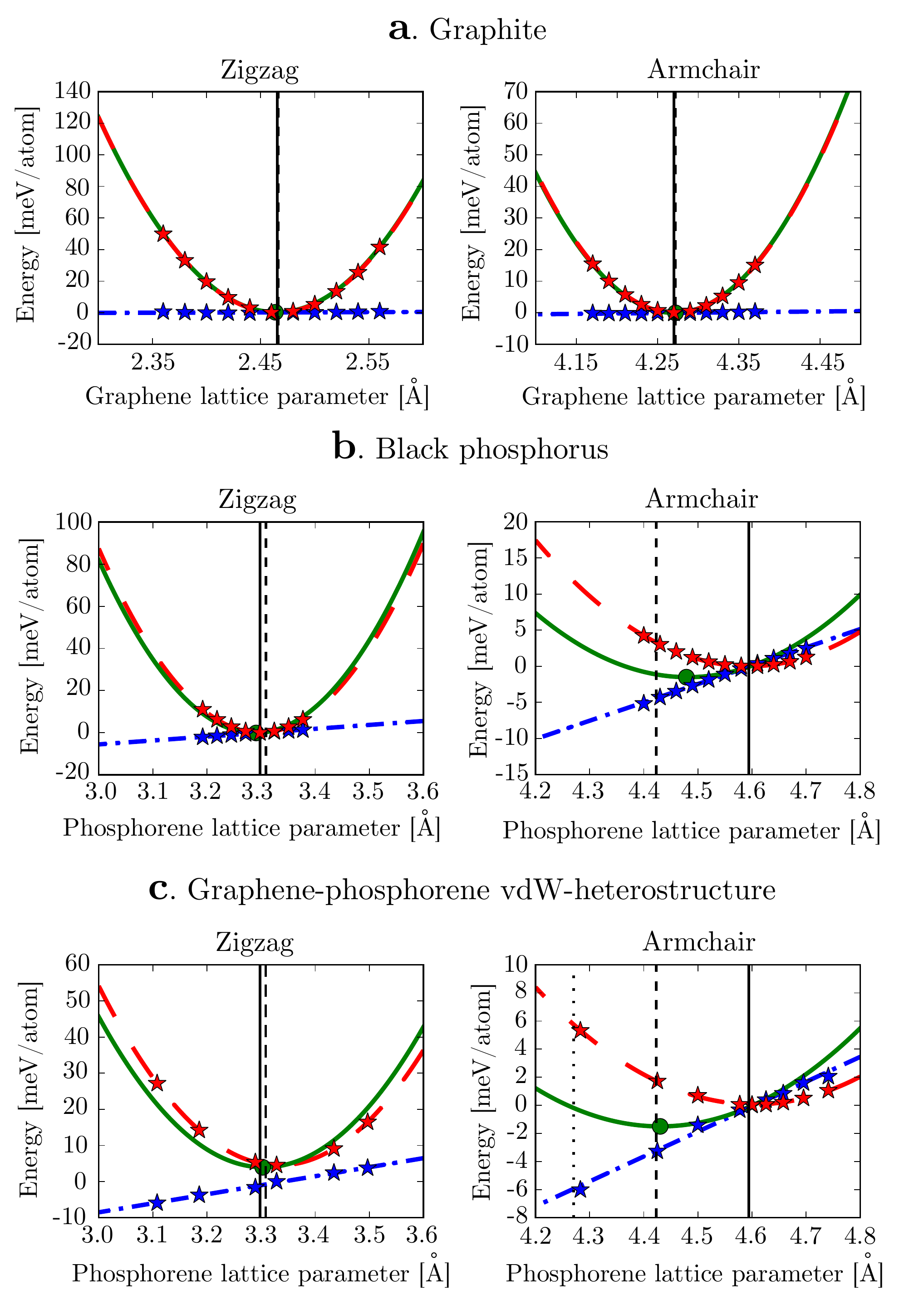}
 \caption{\label{DE_total} Variation of (dashed red) elastic energy, (dotted blue) vdW energy, (solid green) energy sum as defined in Eq.~\ref{DE} as a function of (Left) zigzag lattice parameter;
  (Right) armchair lattice parameter. The top panels correspond to the case of graphite, the middle panels to black phosphorus while the bottom panels correspond to the case of the graphene-phosphorene vdW-heterostructures.
  For comparison, we display in dashed black the corresponding lattice parameter of graphene/phosphorene, in solid black the corresponding lattice parameter of black phosphorus and in dotted black either 4/3 of the zigzag
  lattice parameter of graphene, or directly the armchair lattice parameter of graphene. Stars correspond to the computed values, and circles correspond to the minima of the total energy.}
\end{center}
\end{figure}

The results of the present model are depicted in Fig.~\ref{DE_total} for all the considered systems: on the left, the variation of the vdW and elastic energies are shown with respect to the zigzag lattice parameter of
either graphene (in graphite) or phosphorene (in black phosphorus and in the vdW-heterostructure),
while on the right the same quantities are presented with respect to armchair lattice parameter. Note the change of energy scale between the different panels.
The results for graphite are shown in the top panels (Fig.~\ref{DE_total}a),
black phosphorus in the middle panels (Fig.~\ref{DE_total}b) and the vdW-heterostructure in the bottom panels (Fig.~\ref{DE_total}c).

The change in vdW energy is rather linear with respect to the in-plane lattice parameter, being directly proportional to the density of atoms in the changed layer: the highest the density, the lowest the vdW energy. 
The gain in vdW energy by compressing graphene or phosphorene in the zigzag direction is
nearly negligible when compared to the corresponding loss in elastic energy. In consequence, the relaxed zigzag lattice parameters of graphene and phosphorene in graphite, black phosphorus and
in the vdW-heterostructure are close to the ones of the corresponding free-standing counterparts,
in agreement with what has been reported in Tabs.~\ref{Tab1} and~\ref{Tab1b}. On contrary, in the armchair direction, the gradient of vdW energy plays an important role for phosphorene, 
as it is of the same order of magnitude as the variation in elastic energy. 
Our computations show that, when they are stacked upon another, phosphorene layers will tend to contract in order to decrease their total energy, as observed previously in black phosphorus.
With the simple model presented in this section, we are able to understand the origin of such phenomenon, albeit qualitatively. 
An accurate quantitative description requires the addition of the interlayer change in kinetic, Hartree, Ewald, (non-)local pseudopotential,
exchange-correlation and core pseudopotential energies with strains, which would be the topic of further work. 
Still, our model allows us to understand why the most favorable predicted structure exhibits
a large armchair contraction of phosphorene.

In Fig.~\ref{DE_total}c are presented the results related to the graphene-phosphorene vdW-heterostructure with respect to both zigzag and armchair directions.
The change with graphene lattice parameters is not shown, because it is extremely similar to the graphite case i.e. negligible.
Similarly to black phosphorus, the zigzag parameter of phosphorene in the heterostructure is only weakly affected by the vdW interactions with the graphene layers.
The armchair lattice parameter is on contrary strongly modified by these interactions, and is predicted to contract more than in black phosphorus.
This global $\sim 4 \%$ contraction is predicted independently of the nature of the interface, and thus it should even be present in the case of an incommensurate graphene-phosphorene vdW-heterostructure.
Still, this value remains mostly indicative, as not only our model does not include all the DFT ingredients as it normally should,
but the functional and dispersion corrections have as well their own limited accuracies~\footnote{PBE-D3 overestimates the experimental lattice 
parameter of black phosphorus by approximatively 1\%, see Tab.~\ref{Tab1}.}. 
Furthermore, local accommodations can induce further global straining of phosphorene, up to a $\sim 7 \%$ compression on average for a pseudo-coherent interface between graphene and phosphorene:
the G$_1$P$_1$($0^\circ$, $0^\circ$) studied in Sec.~\ref{Sec2}.
This commensurate structure consists of 3$\times$1 phosphorene and 4$\times$1 graphene conventional cells and has already been quite extensively studied in the
literature~\cite{Padilha2015,Cai2015} in the case of graphene-supported phosphorene. However, to the best of our knowledge, it has been done by fixing the lattice either to the ones of isolated phosphorene 
or to an arithmetic average of graphene and phosphorene lattice constants,
without considering the differences in elastic constants between graphene and phosphorene. Similar hypotheses are made in the case of h-BN/phosphorene vdW-heterostructures~\cite{Cai2015,Hu2015,Steinkasserer2016},
as graphene and h-BN lattice parameters only differ by $\sim 2\%$. 
We stress out here is that it is expected that phosphorene will contracts largely when stacked on graphene or h-BN. The magnitude of this effect remains into question, as well as the type of 
interface (coherent, semi-coherent, or incoherent) that will be formed.



Still, several additional informations can be extracted from this model. First, and most importantly, as it is so soft in the armchair direction, phosphorene will tend to accommodate to any material 
it is in contact with, e.g. graphene, in order to maximize its vdW energy, and the effect can be quite important ($\sim$ 4\%).
This indicates strong substrate effect on the structural properties of phosphorene, and by extension on its electronic properties, which are extremely sensitive to strain~\cite{Peng2014} (direct to indirect gap
transition). 
This will be discussed in more detail in the next section. This effect may also explain why the positions of the Raman peaks of phosphorene monolayer and multi-layers are experimentally reported to be close in
frequency~\cite{Phaneuf-LHeureux2016} (differences smaller than
5 cm$^{-1}$ between phosphorene and black phosphorus Raman peaks), while one would expect the contrary when referring to the change of the armchair lattice parameter between phosphorene and black phosphorus and the
influence of strain on the Raman peaks of phosphorene~\cite{Fei2014} (of the order of 30 cm$^{-1}$ for the change of lattice parameter from monolayer to the bulk predicted i.e.~$\sim 4\%$).
Indeed, if the substrate induces a contraction of the armchair lattice parameter of phosphorene to a value close to the one of black phosphorus, then one would expect relatively close Raman peak positions~\cite{Phaneuf-LHeureux2016}.
As it will be shown, the changes in electronic properties are also important, nearly turning the direct gap of phosphorene into an indirect one, and thus specific features 
should be observable in the resonant Raman spectroscopy~\cite{Corro2016} of supported/encapsulated phosphorene depending on the supporting or encapsulating material.

%





Second, based on our theoretical model, the energy difference between a given commensurate structure, characterized by its average strain tensors and rotation angles (see Tab.~\ref{Tab2}), 
and the estimated ground state of our model can be estimated. To do so, the average strains of the constituent layers in the vdW-heterostructure, as estimated based on our DFT computations, is fed in Eq.~\ref{DE}.
Compared to the aligned case, the elastic energy includes additionally the contributions due to graphene and phosphorene shear strains.
The change in vdW energy is itself estimated based on its linear interpolation along zigzag and armchair
directions done previously in this section, neglecting thus its change with shear strain. The difference between this resulting (total) energy and the ground-state energy estimated by our model 
is then subtracted to the mixing energy computed using Eq.~\ref{Mixing}. Similar corrections are considered in the case of the substitution energy (Eq.~\ref{Substitution}),
but taking additionally into account of the change of elastic and
vdW energies of graphite and black phosphorus with strain.

Finally, the validity of our model is investigated by comparing the elastic constants computed in DFPT for the vdW-heterostructure
and the theoretical model. Indeed, based on Eq.~\ref{dEtotal}, the elastic constants of the heterostructure can be written as:
\begin{equation}
 c^{\text{HS}}_{\alpha\beta} = \frac{1}{d}\left(\tilde{c}^\text{C}_{\alpha\beta} \left[ 1+\epsilon^{\text{C}}_{\alpha} \right] \left[ 1+\epsilon^{\text{C}}_{\beta} \right] +
 \tilde{c}^\text{P}_{\alpha\beta} \left[ 1+\epsilon^{\text{P}}_{\alpha} \right] \left[ 1+\epsilon^{\text{P}}_{\beta} \right] \right), \label{Cmodel}
\end{equation}
where $d$ is the lattice constant in the out-of-plane direction of the heterostructure. The elastic constants computed on the one hand by DFPT for G$_1$P$_1$($0^\circ$, $0^\circ$, $b_P= b_C$)
and on the other hand with Eq.~\ref{Cmodel} for G$_1$P$_1$($0^\circ$, $0^\circ$, $b_P= b_C$) and for the most stable structure predicted by the model, denoted G$_1$P$_1$($0^\circ$, $0^\circ$, $b_P=1.04\times b_C$),
are reported in Tab.~\ref{Tab4}.

\begin{table}[h]
\caption{\label{Tab4} Elastic constants of the graphene-phosphorene vdW-heterostructure G$_1$P$_1$($0^\circ$, $0^\circ$) predicted either by PBE-D3 within the DFPT formalism (only for the commensurate 
where the armchair lattice parameters of graphene and phosphorene are matched G$_1$P$_1$($0^\circ$, $0^\circ$, $b_P=b_C$)), or using the model based on Eq.~\ref{DE}
and Eq.~\ref{Cmodel}. In this last case, we give the predicted elastic constants for G$_1$P$_1$($0^\circ$, $0^\circ$, $b_P=b_C$) and for the vdW-heterostructure which corresponds to the energy minimum in Fig.~\ref{DE_total},
G$_1$P$_1$($0^\circ$, $0^\circ$, $b_P=1.04 \times b_C$). }
\begin{tabular}{l c c c c c c c c c}
 \hline \hline
 & \multicolumn{9}{c}{Elastic constants [GPa]} \\
  \cline{2-10}
 & $c_{11}$ & $c_{22}$ & $c_{12}$ & $c_{33}$ & $c_{13}$ & $c_{32}$ & $c_{44}$ & $c_{55}$ & $c_{66}$ \\
   \cline{2-10}
 &\multicolumn{9}{l}{GP$_1$($0^\circ$, $0^\circ$, $b_P=b_C$)} \\
  \cline{1-10}

   PBE-D3 & 529 & 427 & 102 & 43 & -7 & -4 & 8 & -1 & 208 \\
   Model & 534 & 434 & 93  \\
   \hline 
   & \multicolumn{9}{l}{GP$_1$($0^\circ$, $0^\circ$, $b_P=1.04 \times b_C$)} \\
     \cline{1-10}
   
  Model & 533 & 434 & 94 \\
\hline \hline
\end{tabular}

\end{table}

Overall, this simple model is able to reproduce within 10 \% the elastic constants of G$_1$P$_1$($0^\circ$, $0^\circ$, $b_P=b_C$). When compared to the G$_1$P$_1$($0^\circ$, $0^\circ$, $b_P=1.04 \times b_C$) case,
G$_1$P$_1$($0^\circ$, $0^\circ$, $b_P=b_C$) has similar in-plane elastic constants. Note the sightly negative value for the $c_{55}$ elastic constant, indicating a structural instability due to a shear deformation
along the plane defined by the out-plane and armchair
lattice vectors. The phosphorene layers sandwiching graphene layers would thus be translated compared to another, leading to a different translational stacking than the one considered in G$_1$P$_1$ up to now.
This may also lead to
variations in term of mixing and substitution energy for the G$_1$P$_1$($0^\circ$, $0^\circ$, $b_P=b_C$) heterostructure, which may lead to its stabilization when compared to the isolated phases of graphite and black phosphorus.
Still, this instability is at the edge of DFT precision, and possibly our numerical accuracy, and its impact expected to be small; it will not thus be studied here and is left for future work.

\section{Electronic properties} \label{Sec4}

\begin{figure}[htp]
 \begin{center}
 \includegraphics[width=0.48\textwidth]{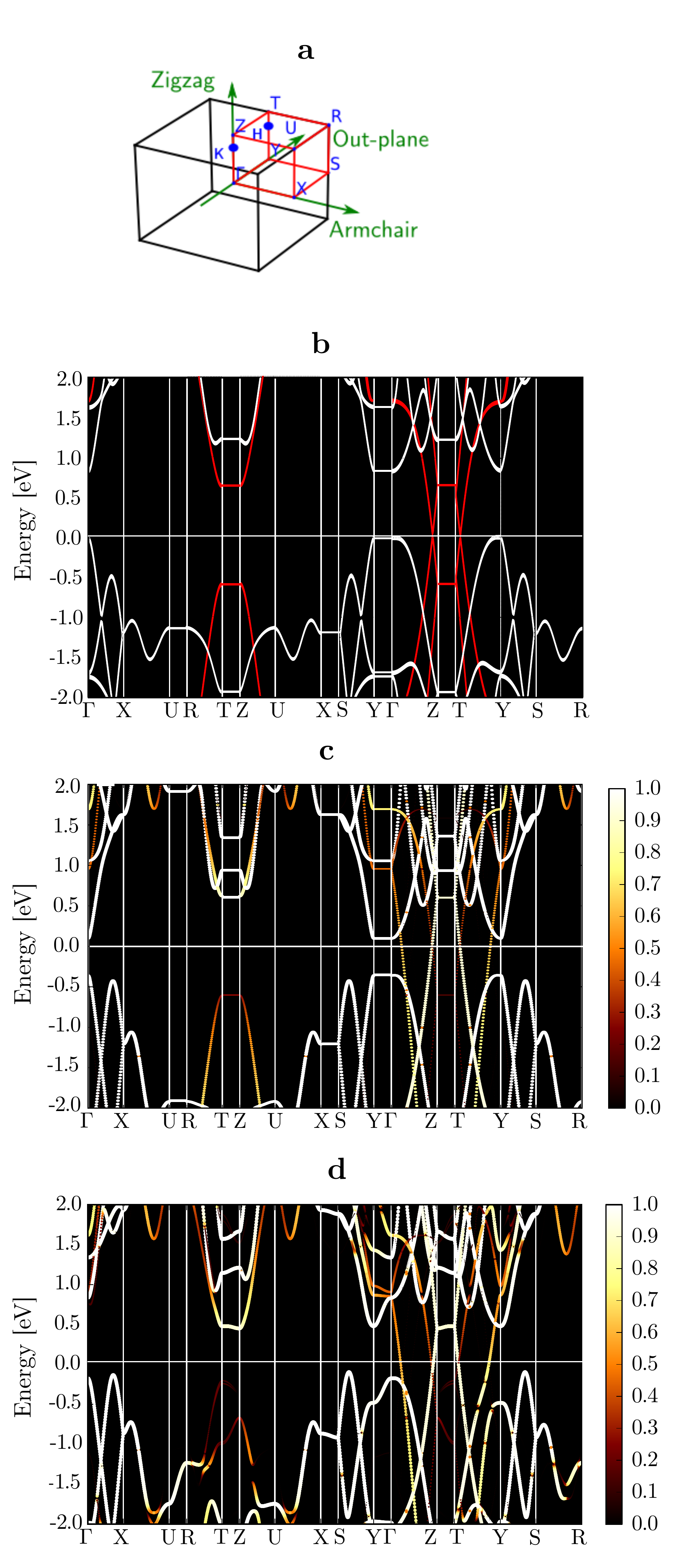}
\caption{\label{Electronic} Electronic band structures, unfolded on the Brillouin zone of phosphorene shown in (a), of (b) phosphorene (in white) and graphene (in red) 
monolayers superposed on top of each other, supposing that the graphene and phosphorene armchair lattices match and that the graphene zigzag lattice parameter is exactly 3/4
of its phosphorene counterpart. (b) graphene and phosphorene strained such
  as the in-plane crystalline structure corresponds to the G$_1$P$_1$($0^\circ$, $0^\circ$, $b_P=b_C$)'s one. The layers are spaced by a 15~\AA-thick vacuum
  layer to avoid in that case the interactions between the layers. (c) G$_1$P$_1$($0^\circ$, $0^\circ$, $b_P=b_C$) graphene-phosphorene vdW-heterostructure.
  The weight associated with the unfolding technique is given by the color bar and the marker size.}
\end{center}
\end{figure}

Following our observations on the intralayer deformation of phosphorene in the graphene-phosphorene vdW-heterostructures, the impact of such deformation on their electronic properties 
is investigated in the present section, and more specifically in the case of G$_1$P$_1$($0^\circ$, $0^\circ$, $b_P=b_C$), taken as an approximation of the ground-state structure 
G$_1$P$_1$($0^\circ$, $0^\circ$, $b_P=1.04 \times b_C$) predicted by the theoretical model of Sec.~\ref{Sec3}.
To do so, the electronic band structure of phosphorene is recalled in Fig.~\ref{Electronic}b, alongside its corresponding Brillouin's zone (Fig.~\ref{Electronic}a). 
On top, the electronic band structure of graphene is superimposed, in its conventional cell, supposing that its armchair reciprocal lattice vector~\footnote{The reciprocal lattice vector parallel
to the armchair crystalline direction of phosphorene.} is matching its phosphorene counterpart,
and that the zigzag reciprocal lattice vector is exactly 3/4 of its phosphorene counterpart. The Fermi levels of graphene and phosphorene have been aligned.
Although fictional -obviously graphene and phosphorene lattices do not match, thus their reciprocal lattices do not match neither-, this approach allows us to see where band crossings could potentially occur. 
Graphene is a zero-gap semi-conductor, while phosphorene is a real semi-conductor, with a nearly-direct gap at $\Gamma$ of 0.84 eV predicted by PBE-D3. This value compares
relatively well with the 1~eV obtained with PBE~\cite{Liu2014} or with the 0.89~eV computed with PBE corrected by Grimme's DFT-D2~\cite{Qiu2017}. 
However, it differs significantly from the experimental values reported up to now (between
1.4 and 2~eV~\cite{Castellanos2015}), due, mainly, to the well-known underestimation of the gap in DFT~\cite{Martin2004}. 
Note that, close to the Fermi level,
the bands of phosphorene and graphene do not cross each other, except along the $R-T$, $Z-U$, $Y-Z$ and $T-Y$ lines.
For sake of comparison, we recall that graphite is a semi-metal~\cite{Terrones2010} and that
black phosphorus is reported experimentally as a direct band gap semiconductor with a gap of $\sim 0.261$~eV at 40~K~\cite{Ehlen2016}.
 Similarly to what is reported for example in Ref.~\onlinecite{Rudenko2014}, but with different dispersion corrections for the long-range e$^-$-e$^-$ correlation, our calculations predict black phosphorus 
to be metallic. In order to recover the correct electronic character of this material, beyond-DFT techniques (like G$_0$W$_0$) should be used~\cite{Rudenko2014}.

Before discussing
the electronic band structure of G$_1$P$_1$($0^\circ$, $0^\circ$, $b_P=b_C$), one has to remind that the electronic properties of phosphorene
are quite sensitive to strain, as reported for example in Ref.~\onlinecite{Peng2014}. 
In order to de-correlate its effects from the one arising from the interactions with graphene (hybridization), we show the electronic band structure of graphene and phosphorene in Fig.~\ref{Electronic}c
using the G$_1$P$_1$($0^\circ$, $0^\circ$, $b_P=b_C$) in-plane geometry, but separating the layers by a 15~\AA-thick vacuum layer such as the (electronic) interactions
between the layers can be neglected. They are unfolded on the primitive cell of phosphorene using \textsc{fold2bloch}~\cite{Rubel2014} as available in the \textsc{abinit} software.
Note that the strain in the armchair direction ($\sim$ 7\%) can lead to important changes in the electronic band structures of
the constituent layers.
Notably, the second maximum in the valence band of phosphorene
along the $\Gamma-X$ high symmetry line is shifted upwards, and the minimum of the conduction band in phosphorene is strongly shifted downwards, shrinking its electronic gap by $~\sim 0.35$~eV,
which is consistent with the observations in Ref.~\onlinecite{Peng2014}. Graphene
is only weakly affected by strain.

The electronic properties of the G$_1$P$_1$($0^\circ$, $0^\circ$, $b_P=b_C$) structure, including thus now the interlayer interactions, is shown in Fig.~\ref{Electronic}d. 
Compared to the previous case, where only the effects of strains were considered, several additional changes are noticeable. 
First, at the crossing of graphene and phosphorene bands in the lattice-matched case, avoided crossings can be observed in the electronic band structure of the vdW-heterostructure,
except along $\Gamma-Z$ where the linear band
of graphene crosses a band of phosphorene without being affected. We suppose in that case that the states are simply orthogonal to another.
Second, the strain effect on the conduction band minimum of phosphorene is counterbalanced by the interaction with graphene, leading to
a comparable direct electronic transition at $\Gamma$ for phosphorene free-standing or in the vdW-heterostructure. This effect can announce a quantum confinement in the monolayer, 
that disappears moving to the bulk graphene-phosphorene vdW-heterostructure.
Third, the band offset is quite small ($<$0.1 eV), indicating
no important charge transfer between graphene and phosphorene layers. Fourth, large modifications of the band structures are observed out-of-plane, where lifts of degeneracy of the graphene and 
phosphorene electronic states are observed. 
The G$_1$P$_1$($0^\circ$, $0^\circ$, $b_P=b_C$) vdW-heterostructure is semi-metallic, similarly to graphite. 
The first electronic transition between phosphorene states is found be indirect.

In consequence, the intrinsic electronic properties of phosphorene can be tuned by the environment (substrate or other), not only by state-hybridization or dielectric screening,
but also by vdW-induced strain as discussed in this paper. This finding may reveal to be critical for electronic applications, where it has been proposed to encapsulate phosphorene in h-BN to avoid its oxidation~\cite{Favron2015,Constantinescu2016,Qiu2017} without
altering its electronic band structure. While the effect of state-hybridization~\cite{Constantinescu2016} and dielectric screening~\cite{Qiu2017} have already been investigated, the effect
of vdW-induced strain should also play an important role, and should be investigated in more details for the h-BN/phosphorene/h-BN vdW-heterostructure. Depending on the supporting or encapsulating material,
it may even be possible to close the gap of phosphorene using such vdW-induced strain, although the strain required may be too important (contraction $>$12\%, according to Ref.~\onlinecite{Peng2014})
to be achievable by vdW interactions only.


\section{Conclusion}

In the present work, the properties of bulk graphene-phosphorene vdW-heterostructures have been investigated. First, their stability has been analyzed as a function of the rotation angle between the layers for
different definition of chemical potentials and as a function
of the stacking composition, thus finding
that the most stable configuration corresponds to the alignment of graphene and phosphorene crystalline directions. A new concept of substitution energy is proposed to investigate
this specific degree of freedom. The accommodation of phosphorene to graphene in the vdW-heterostructure is then examined, notably in the armchair direction. To do so, a model based on the elastic and vdW energies
is constructed, explaining the contraction
of phosphorene armchair lattice parameter from the monolayer to the bulk (black phosphorus).
In the graphene-phosphorene vdW-heterostructures, it is found that phosphorene contracts as well in order to maximize its vdW interactions with graphene, leading to strong structural and electronic
modifications. This effect is identified as independent of the nature of the interfaces (coherent, semi-coherent or incoherent). This leads to the conclusion that, 
in average, phosphorene will always compress quite importantly compared to its free-standing form in order to maximize its interactions with any material it is in contact with,
even if they are bound only by weak dispersive forces. Similar effects are expected in other soft 2D materials, and the present model can be easily transferred to study interfaces based on such materials, and 
further refined in order to include the effects of local accommodations, that may allow to predict the formation of semi-coherent or incoherent interfaces.

In consequence, this work calls into questions the role of the substrate on the intrinsic properties of phosphorene; such role has to the best of our knowledge always been neglected up to now and thus may
be re-investigated. More specifically, the impact of encapsulation of phosphorene between h-BN layers may lead to 
either undesired changes in the intrinsic electronic properties of phosphorene
(direct to indirect gap transition) or on contrary positive changes, like an increase of the electronic mobility as observed for example in graphene encapsulated between
h-BN layers~\cite{Dean2010}. We also mention
the possibility of substrate-controlled stress-tronic in phosphorene, where the gap could be tuned only based on the substrate.
In parallel, the investigation of the sodiated phases of these graphene-phosphorene vdW-heterostructures
may allow to understand better why their performance as anode in sodium-ion batteries are
improved comparatively to black phosphorus or graphite~\cite{Sun2015}. 

\section*{Acknowledgments}

The authors acknowledge technical help from J.-M.~Beuken and M. Giantomassi and scientific discussions with S. M.-M. Dubois and D. Di Stefano. This work has been supported by the FRS-FNRS through a FRIA
Grant (B.V.T.) and a research project (N$^\circ$T.1077.15); the Communauté française de Belgique through the BATTAB project (ARC 14/19-057) and
the \textit{``3D nanoarchitecturing of 2D crystals''} project (ARC 16/21-077); the Région Wallone through the BATWAL project (N$^\circ$1318146);
the European Union's Horizon 2020 research and innovation program (N$^\circ$696656). 
Computational resources have been provided by the supercomputing facilities of
the Université catholique de Louvain (CISM/UCL) and the Consortium des
Equipements de Calcul Intensif en Fédération Wallonie Bruxelles (CECI) funded by
the Fonds de la Recherche Scientifique de Belgique (FRS-FNRS) under convention 2.5020.11.
The present research benefited from computational resources made available on the Tier-1 supercomputer of the Fédération Wallonie-Bruxelles, 
infrastructure funded by the Walloon Region under the grant agreement N$^\circ$1117545.

\end{document}